\documentclass[prd,showpacs,twocolumn,superscriptaddress,floatfix,10pt]{revtex4-2}
\usepackage{setspace} 
\usepackage[utf8]{inputenc}
\usepackage{float}
\usepackage{amsmath,amssymb,amsfonts,bm}
\usepackage{graphicx}
\usepackage{multirow}
\usepackage[usenames,dvipsnames]{color}
\allowdisplaybreaks
\usepackage[
colorlinks=true,
linkcolor=blue,
breaklinks=true,
urlcolor=blue,
citecolor=blue]{hyperref}

\usepackage{orcidlink}

\definecolor{green}{rgb}{0,0.6,0}

\newcommand{\lhc}{{\rm lhc}}
\newcommand{\rhc}{{\rm rhc}}

\newcommand{\be}{\begin{equation}} 
\newcommand{\ee}{\end{equation}}
\newcommand{\bea}{\begin{eqnarray}} 
\newcommand{\eea}{\end{eqnarray}}
\newcommand{\beas}{\begin{eqnarray*}} 
\newcommand{\eeas}{\end{eqnarray*}}

\newcommand{\veq}{{\bm q}}
\newcommand{\vek}{{\bm k}}
\renewcommand{\vec}{\bm}
\newcommand{\Tcc}{T_{cc}^+}

\newcommand{\uestc}{\affiliation{School of Physics, University of Electronic Science and Technology of China, Chengdu 611731, China}}

\newcommand{\ific}{\affiliation{Instituto de F\'isica Corpuscular (centro mixto CSIC-UV), \\
Institutos de Investigaci\'on de Paterna, Apartado 22085, 46071, Valencia, Spain}}

\newcommand{\rub}{\affiliation{Institut f\"ur Theoretische Physik II, Ruhr-Universit\"at Bochum, D-44780 Bochum, Germany }}

\newcommand{\fzj}{\affiliation{Institute for Advanced Simulation, Institut f\"ur Kernphysik and J\"ulich Center for Hadron Physics, Forschungszentrum J\"ulich, D-52425 J\"ulich, Germany}}

\newcommand{\itp}{\affiliation{CAS Key Laboratory of Theoretical Physics, Institute of Theoretical Physics, \\Chinese Academy of Sciences, Beijing 100190, China}}

\newcommand{\ucas}{\affiliation{School of Physical Sciences, University of Chinese Academy of Sciences, Beijing 100049, China}}

\newcommand{\scnu}{\affiliation{Guangdong Provincial Key Laboratory of Nuclear Science, Institute of Quantum Matter, South China Normal University, Guangzhou 510006, China}}

\newcommand{\ihep}{\affiliation{Institute of High Energy Physics, Chinese Academy of Sciences, Beijing 100049, China}}

\newcommand{\qmscnu}{\affiliation{Guangdong-Hong Kong Joint Laboratory of Quantum Matter, Southern Nuclear Science Computing Center, South China Normal University, Guangzhou 510006, China}}

\newcommand{\JSI}{\affiliation{Jozef Stefan Institute, Jamova 39, 1000 Ljubljana, Slovenia}}

\newcommand{\lis}{\affiliation{CeFEMA, Center of Physics and Engineering of Advanced Materials, Instituto Superior T{\'e}cnico, Avenida Rovisco Pais 1, 1049-001 Lisboa, Portugal}}

\newcommand{\peng}{\affiliation{Peng Huanwu Collaborative Center for Research and Education, Beihang University, Beijing 100191, China}}

\begin{document}
\title{Role of left-hand cut contributions on pole extractions from lattice data:\\
Case study for $T_{cc}(3875)^+$}

\begin{abstract}
We discuss recent lattice data for the $T_{cc}(3875)^+$ state to stress, for the first time, a
potentially strong impact of left-hand cuts from the one-pion exchange on the pole extraction for near-threshold exotic states. In particular, if the left-hand cut is located close
to the two-particle threshold, which happens naturally in the $DD^*$ system for the pion mass exceeding its physical value, the effective-range expansion is valid only in a very limited energy range up to the cut and as such is of little use to reliably extract the poles. Then, an accurate extraction of the pole locations requires the one-pion exchange to be implemented explicitly into the scattering amplitudes. Our findings are general and potentially relevant for a wide class of hadronic near-threshold states.
\end{abstract}

\author{Meng-Lin Du\orcidlink{0000-0002-7504-3107}}
\uestc

\author{Arseniy~Filin\orcidlink{0000-0002-7603-451X}}
\rub

\author{Vadim Baru\orcidlink{0000-0001-6472-1008}}
\rub 

\author{Xiang-Kun Dong\orcidlink{0000-0001-6392-7143}}
\itp\ucas 

\author{Evgeny Epelbaum\orcidlink{0000-0002-7613-0210}}
\rub

\author{Feng-Kun~Guo\orcidlink{0000-0002-2919-2064}}
\itp \ucas \peng

\author{Christoph~Hanhart\orcidlink{0000-0002-3509-2473}}
\fzj 

\author{Alexey Nefediev\orcidlink{0000-0002-9988-9430}}
\JSI
\lis

\author{Juan Nieves\orcidlink{0000-0002-2518-4606}}
\ific

\author{Qian Wang\orcidlink{0000-0002-2447-111X}}
\scnu\ihep\qmscnu

\maketitle

{\it Introduction.}---The last two decades have witnessed the discovery of a large number of the so-called exotic hadronic states
in the heavy quark sector that do not fit into the scheme of simple quark models~\cite{Esposito:2014rxa,Lebed:2016hpi,Chen:2016qju,Guo:2017jvc,Kalashnikova:2018vkv,Yamaguchi:2019vea,Brambilla:2019esw,Guo:2019twa,Chen:2022asf}. For some of them only particular properties like masses or decays strongly deviate from expectations, for others already the quantum numbers unambiguously
indicate their multiquark content---the most prominent representatives of this class are 
the isotriplet $Z_c$ and $Z_b$ states that decay to a heavy quarkonium plus a single pion. For an overview of the experimental situation see, {\it e.g.},~\cite{Brambilla:2019esw}. 

The pressing theoretical question is what clusters the quarks form in these exotic hadrons.
One popular scenario is that diquarks and anti-diquarks emerge as relatively compact
building blocks carrying a color charge~\cite{Esposito:2014rxa,Lebed:2016hpi}.
Then, if there are light quarks in the system, the size of the emerging states is governed by the confinement radius, $\sim1/\Lambda_\text{QCD}$ ($\Lambda_\text{QCD}\sim 300$~MeV), and does not depend on the binding energy $E_b$ defined as the difference between the mass of the state and the energy of the closest threshold.
Alternatively, the building blocks could be color-neutral conventional 
hadrons. Then, the size of these so-called hadronic molecules is given by 
the inverse of the binding momentum 
$\gamma=\sqrt{2\mu E_b}\ll\Lambda_{\rm QCD}$, with $\mu$ for the reduced mass, 
resulting in large radii of near-threshold states. Thus, the size of hadronic molecules is limited by the binding momentum rather than by the structure of the interaction. This difference in
size leads to significant differences in some observables~\cite{Guo:2017jvc}. 
Because of confinement, only color singlet multi-hadron intermediate states can go on-shell thereby generating a unitarity or right-hand cut (rhc) in the amplitude. 
The argument can be further generalized to unstable
constituents~\cite{Braaten:2007dw,Hanhart:2010wh}, as long as they are not too broad~\cite{Filin:2010se}, with lifetime larger than the range of forces~\cite{Guo:2011dd}. 

So far little is known about the  potential that leads to exotics---this changes with identifying the potentially
strong impact of left-hand cuts (lhcs) from the one-pion exchange on the pole extraction for doubly heavy near-threshold
states, which is demonstrated here for the first time. For concreteness, we present our
findings with the focus on the 
double-charm meson $\Tcc$ reported in \cite{LHCb:2021vvq,LHCb:2021auc} and treated as a $DD^*$ molecule. Among many exotic candidates, the $\Tcc$ is of particular interest since its width, apart from a tiny electromagnetic contribution,
stems almost entirely from the only available strong decay channel $DD\pi$. Then, 
the only relevant cut of the amplitude on the real axis is the three-body $DD\pi$ cut (blue dashed vertical lines in Fig.~\ref{fig:cuts}), while the $DD^*$ branch cut (green dashed vertical line) splits into a pair of cuts on the second sheet in the complex energy plane \cite{BGMT,Doring:2009yv}. The branch point of the $DD\pi$ cut is located below the nominal $DD^*$ threshold. 

\begin{figure}[t]
\begin{center}
\includegraphics[width=0.48\textwidth]{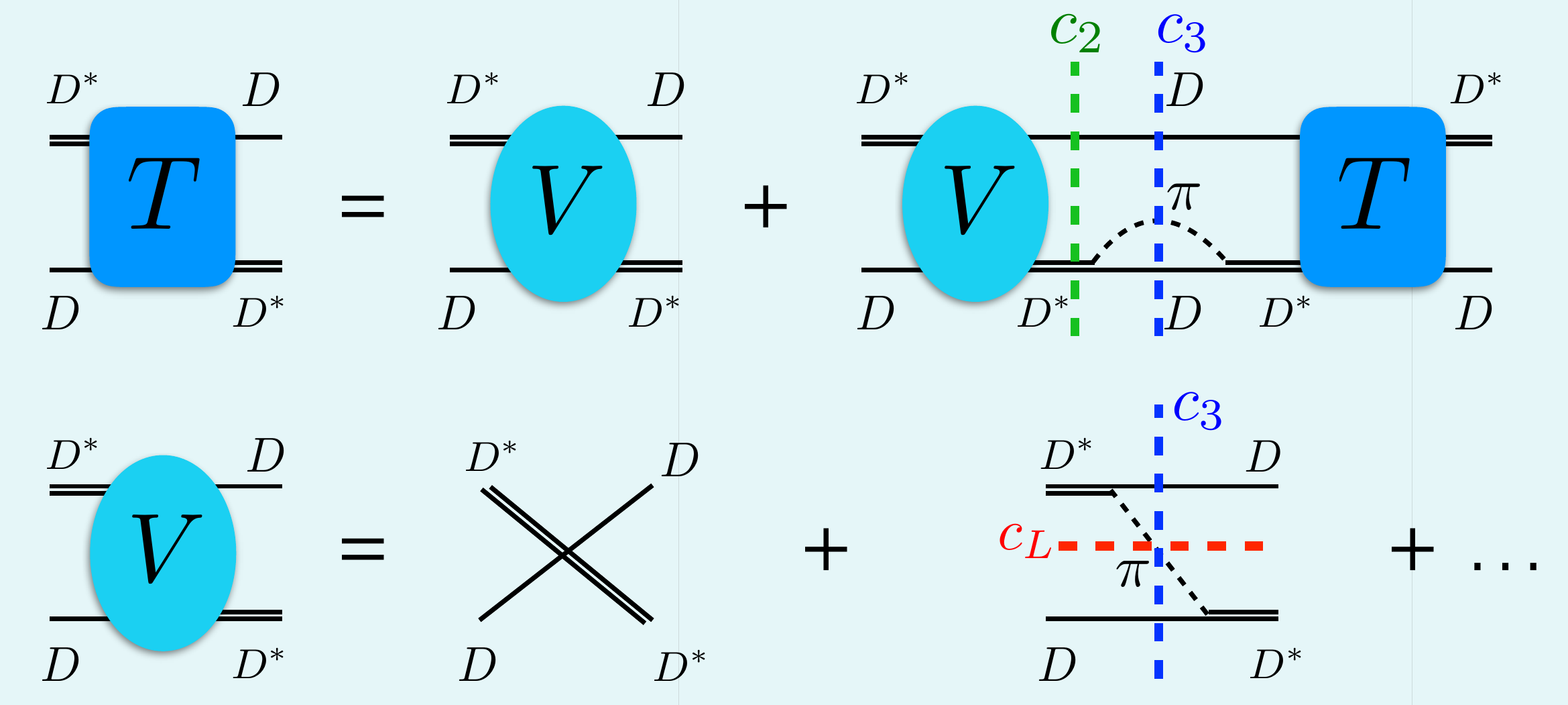}
\caption{The cut structure in the $DD^*$ system: (i) the blue dotted vertical 
lines ($c_3$) indicate the three-body right-hand cuts, (ii) the green dotted vertical line ($c_2$) shows the two-body cut, and (iii) the red dotted horizontal line ($c_{\rm L}$) is for
the left-hand cut. $T$ and $V$ denote the amplitude and interaction potential, respectively.
}
\label{fig:cuts}
\end{center}
\end{figure}

The cut structure of the amplitude severely changes for a heavier pion, which can be studied using lattice QCD and chiral effective field theories. As soon as the pion mass $m_\pi$ exceeds the $D^*$-$D$ mass difference $\Delta M=M_{D^*}-M_D$, the $\Tcc$ is stable with respect to the strong interaction, the $DD\pi$ three-body cut branch point appears above the $DD^*$ two-body threshold, and the pion exchange induces the lhc (the red dashed horizontal line in Fig.~\ref{fig:cuts})~\cite{Frautschi:1960qzm,Oller:2019rej}. Since the location of the branch point is related to the range of the potential and the discontinuity depends on its strength, the lhc is also called dynamical~\cite{omnes1971}. Other relevant cuts present in the system are discussed below.

The $\Tcc$ was recently studied in lattice QCD \cite{Padmanath:2022cvl,Chen:2022vpo,Lyu:2023xro}. The last work employs the HAL~QCD method to extract the $DD^*$ scattering potential and then use it to calculate the phase shifts above the two-body threshold. Our consideration does not directly apply to that approach. In the first two works the L\"uscher method is employed to extract the $DD^*$ phase shifts $\delta(E)$ at $m_\pi=280$ and 350~MeV, respectively. However, in \cite{Chen:2022vpo}, there is only one data point in the near-threshold region and only a single lattice volume is investigated, so the authors themselves argue that a proper discussion of the pole structure of the amplitude is not possible yet. Therefore, we stick to the phase shifts extracted in \cite{Padmanath:2022cvl} and related to the scattering $T$-matrix as
\begin{eqnarray}
-\frac{2\pi}{\mu}T^{-1}(E)=p \cot \delta-ip,
\label{phvsT}
\end{eqnarray}
with $E$ and $p$ the energy and the magnitude of the relative momentum in the center-of-mass (c.m.) frame, respectively. A pole of the $T$-matrix appears if 
\begin{equation}
p \cot\delta = ip.
 \label{poledef}
\end{equation}
 
To exploit condition (\ref{poledef}), the phase shifts extracted from the lattice calculations were parameterized retaining the first two terms in the effective range expansion (ERE)~\footnote{If one of the scattering particles is unstable, it is convenient to make ERE 
around the complex branch point connected to the two-body channels~\cite{Braaten:2009jke,Baru:2021ldu}.},
\begin{equation}
 p \cot\delta = \frac{1}{a} + \frac12 r p^2 + \mathcal{O}(p^4),
 \label{EREdef}
\end{equation}
where $a$ and $r$ are the scattering length and effective range, respectively. However, the convergence radius of ERE is set by the location of the nearest singularity irrespective of its origin. We argue that, in the settings of~\cite{Padmanath:2022cvl}, the physics related to the lhc is relevant and cannot be ignored. In particular, we demonstrate that the simple approximation (\ref{EREdef}) has to be abandoned in favor of the exact solution of the dynamical equation in the presence of pions, which have a strong effect on both $p \cot\delta$ and the extracted pole. Thus, the physics discussed in this Letter is related to the lhcs from long-range potentials and is quite general. 

Importantly, the phase shifts extracted from the lattice data and employed in our analysis may need to be revisited since the presence of the lhc requires a modification of the L\"uscher method \cite{Raposo:2023nex,Dawid:2023jrj} and may induce sizable partial-wave mixing effects~\cite{Meng:2021uhz}.
Being unable to assess quantitatively the importance of these effects, we take the phase shifts extracted above the lhc for granted.

{\it Cut structure of the $DD^*$ amplitude.}---In line with the lattice setting employed to analyze the $DD^*$ system, we work in
the isospin limit and use isospin-averaged masses for all mesons. Then the $T_{cc}(3875)^+$ is a purely isoscalar state. The relevant degrees of freedom are $DD^*$ and $DD\pi$, introducing two- and three-body branch points
in the amplitude, respectively. For the physical pion mass, pions contribute to the $DD^*$ dynamics in two ways: through the $D^*$ selfenergy and $DD^*$ scattering potential. Both induce 
rhcs to the amplitude (upper and lower blue dashed vertical lines in Fig.~\ref{fig:cuts}, respectively). The three-body $DD\pi$ Green's function generating the most relevant cut in the $DD^*$ scattering amplitude 
reads~\cite{Baru:2011rs,Schmidt:2018vvl,Braaten:2020nmc,Du:2021zzh} 
(for details on time-ordered perturbation theory see, {\it e.g.}, \cite{Schweber:1961zz,Baru:2019ndr,supp})
\begin{equation}
\begin{split}
G_\pi^{-1}(E,\vec k\, ',\vec k) &=E-E_D(k^2)-E_D(k^{\prime 2})-\omega_\pi(q^2)\\
&\approx E-2M_D-\frac{k^2+k'^2}{2M_D}-\omega_\pi(q^2) \, ,
\end{split}
\end{equation}
where $E$ is the total energy, $\vec k$ ($\vec k'$) is the incoming (outgoing) $DD^*$ relative momentum
in the c.m. frame, $E_D(k)=\sqrt{M_D^2+k^2}$, $\omega_\pi(q^2)=\sqrt{m_\pi^2+q^2}$,
with $\veq=\vec k-\vec k'$, and 
$k^{(\prime)}=|\vek^{(\prime)}|$, $q=|\veq|$. 

For the physical pion mass, $m_\pi<\Delta M$, and one can find real values
of $k$ and $k'$ such that $G_\pi^{-1}=0$ for each energy $E$ above the three-body threshold $E_{\rm thr}\equiv 2M_D+m_\pi$, with the interpretation of the $DD\pi$ state going on-shell. At the same time, the $DD^*$ Green's function has two branch points in the complex $s$ plane ($s=E^2$), giving rise to the two-body unitarity cuts in the scattering amplitude---see Fig.~\ref{fig:polesandcuts}~(b).

If $m_\pi$ increases, $\Delta M$, governed by heavy-quark spin symmetry violation and not chiral dynamics, changes very little, but the phase space for the $DD\pi$ state shrinks. For $m_\pi>\Delta M$, $D^*$ becomes
stable (its radiative decays are not considered in lattice calculations, so we neglect them too), and the $DD^*$ branch cut moves to the real axis. We measure the energy relative to the $DD^*$ threshold to write $E= M_{D^*}+M_D+p^2/(2\mu)$, where $\mu= M_{D^*}M_{D}/(M_{D^*}+M_{D})$. The two-body branch point is located at $E=M_D+M_{D^*}$, which implies
\begin{equation}
p_{\rhc_2}^2=0.
\end{equation}

Then,
\begin{equation}\label{Eq:Gpi2}
G_\pi^{-1}(E,\vek',\vek) = \Delta M+\frac{p^2}{2\mu}-\frac{k^2+k^{\prime^2}}{2M_D}-\omega_\pi(q^2),
\end{equation}
so the three-body branch point (we set $k=k'=0$) is
\begin{equation}
p_{\rhc_3}^2 = 2\mu(m_\pi-\Delta M).
\label{prhc3}
\end{equation}

In addition, new singularities 
emerge from \eqref{Eq:Gpi2} in the 
on-shell partial-wave amplitudes 
at imaginary values of the momenta, $k^2=k^{\prime2}=p^2<0$. The smallest in magnitude value of $p^2$ where this happens ($\mu\approx M_D/2$) is given by $\omega_\pi(4p^2)\approx\Delta M$ (backward
scattering), 
\begin{equation}
(p_\lhc^{1\pi})^2\approx\frac14[(\Delta M)^2-m_\pi^2].
\label{plhc}
\end{equation}
This sets the location of the branch point for the lhc nearest to the threshold. The other, remote, end-point of this lhc is set by forward scattering~\cite{supp}. The cut structure in the complex $s$ plane for $m_\pi>\Delta M$ (stable $D^*$) is shown in Fig.~\ref{fig:polesandcuts}~(a)
with the lhc located below the two-body $DD^*$ threshold. Decreasing $m_\pi$, the lhc shrinks and moves towards the threshold. For $m_\pi=\Delta M$ both branch points of the lhc and 3-body rhc (see (\ref{prhc3})) hit the threshold. After that, as the $D^*$ becomes unstable, the only relevant cut left on the real axis is the 3-body $DD\pi$ cut---see Fig.~\ref{fig:polesandcuts}~(b). Meanwhile, the other possible lhcs from the two-pion or heavier meson exchanges may still be present. Those are much further away from the threshold and irrelevant to the discussion here.

A lhc introduces nonanalyticity to $p \cot\delta$ defined in \eqref{phvsT}
and, accordingly, sets the upper bound on the convergence radius of ERE \eqref{EREdef}. 
For the case at hand, $p \cot\delta$ acquires an imaginary part for energies below the lhc and cannot be treated as real-valued. 

\begin{figure}[t!]
\begin{center}
\includegraphics[angle=0,width=\linewidth]{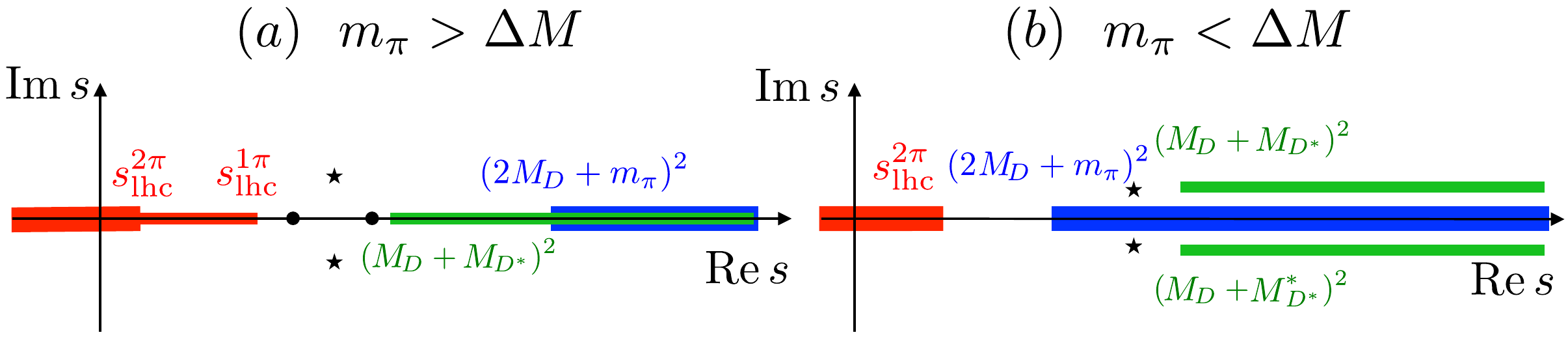}
\caption{Sketch of the locations of various branch cuts and poles in the complex $s$-plane for: (a)
$m_\pi=280$ MeV and (b) the physical 
pion mass. The left-hand, two-body $DD^*$, and three-body $DD\pi$ cuts are shown in red, green, and blue, respectively. The black symbols show typical locations for the $\Tcc$ poles: 
they can correspond to a pair of virtual states (dots) or a resonance (stars) in case (a) and to a quasi-bound state, which would be a bound state for stable $D^*$, in case (b). In cases (a) and (b), the poles are on the second and first Riemann sheets, respectively, with respect to the $DD^*$ cut.}
\label{fig:polesandcuts}
\end{center}
\end{figure}

\begin{figure*}[t!]
\begin{center}
\includegraphics[width=0.9\textwidth]{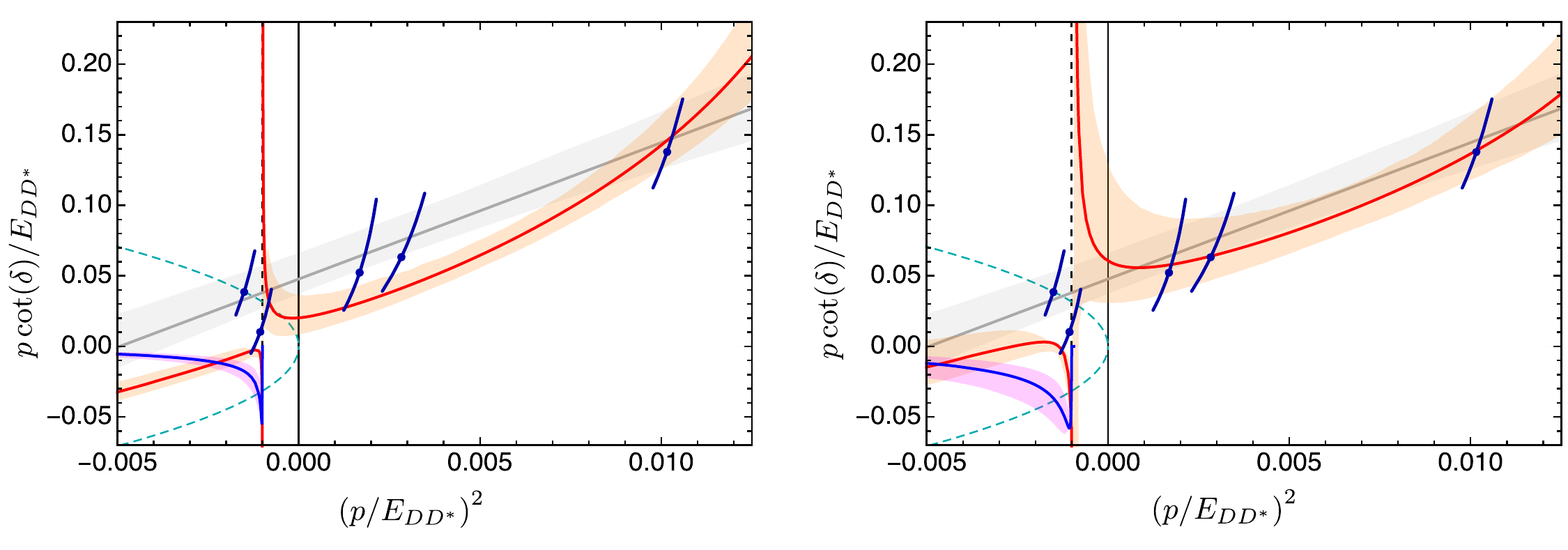}
\caption{Fit results for the lattice data from \cite{Padmanath:2022cvl}. 
The solid and dashed vertical lines indicate the $DD^*$ threshold and the lhc, respectively, while the green dashed curve shows the function $ip$. The gray line and gray band show the fit and its 1$\sigma$ uncertainty, respectively, found in \cite{Padmanath:2022cvl} using ERE formula from (\ref{EREdef}) in the entire energy range both above and below lhc.
The red solid line and the orange band show the best fit and its 1$\sigma$ uncertainty, respectively, calculated in this work. In the right panel, only the three most right data points were used in the fit while in the left panel, in addition, a part of the error bar of the fourth (from right to left) point located above the lhc (for $(p/E_{DD^*})^2>-0.0010$) is included in the fit for illustrative purpose. Below the lhc, $p \cot\delta$ acquires an imaginary part, which is shown as the blue line with the pink uncertainty band.
}
\label{fig:fit}
\end{center}
\end{figure*}

{\it Results and discussion.}---For an illustration of the general concept, we focus on the lattice data from \cite{Padmanath:2022cvl} 
collected at $m_\pi=280$~MeV, $M_D=1927$~MeV and $M_{D^*}=2049$~MeV (the second data set provided in \cite{Padmanath:2022cvl} leads to analogous results~\cite{supp}). Then, with $\Delta M=122~\mbox{MeV}<m_\pi$, we find 
\begin{equation}
(p_\lhc^{1\pi})^2=-(126 \ \mbox{MeV})^2\Longrightarrow \left(\frac{p_\lhc^{1\pi}}{E_{DD^*}}\right)^2=-0.0010,
\end{equation}
where $E_{DD^*}=M_D+M_{D^*}$. In Fig.~\ref{fig:fit} the location of this lhc branch point is indicated with the dashed vertical line. The $DD\pi$ rhc branch point is located far away, 
\begin{equation}
p_{\rhc_3}^2=(552 \ \mbox{MeV})^2 \Longrightarrow \left(\frac{p_{\rhc_3}}{E_{DD^*}}\right)^2=+0.019,
\label{E3}
\end{equation}
and is irrelevant for the current analysis.

In \cite{Padmanath:2022cvl} the $T$-matrix pole is extracted from the lattice data using a linear (in $p^2$) fit, as defined in \eqref{EREdef}, in the full energy range, thereby yielding a pole location at $p_0^2/(2\mu)=-9.9$ MeV [$(p_0/E_{DD^*})^2=-0.0012$], on the second Riemann sheet of the complex $s$ plane. In Fig.~\ref{fig:fit} this fit is shown as the gray line including its 1$\sigma$ uncertainty band in light gray. 
Thus, the pole extracted using only the first two terms in ERE is located below the lhc. However, since the lhc sets the radius of convergence of ERE, the latter is only valid in a small range $|(p/E_{DD^*})^2| < 0.0010$, where no lattice data exist. Furthermore, the central values of the two data points crucial for the fit lie below the lhc, though $p \cot\delta$ should be complex in this case. Therefore, this pole extraction procedure cannot be regarded as reliable.

To improve on the extraction of the pole parameters from the phase shifts
we fit the lattice data from \cite{Padmanath:2022cvl} with an amplitude that includes the lhc~\footnote{An important role played by the lhc from pion exchanges for $NN$ scattering at unphysical pion masses was pointed out in \cite{Baru:2015ira,Baru:2016evv}.}. More precisely, we solve the $DD^*$ scattering equation employing a potential that incorporates
the OPE and two contact 
terms (one momentum-independent and one momentum-dependent) treated as fitting parameters. This is a simplified version of the full amplitude of \cite{Du:2021zzh}, where $D$ waves are now
switched off to be in line with \cite{Padmanath:2022cvl}.
The chiral extrapolation of the pion decay constant is considered using chiral perturbation theory~\cite{Gasser:1983yg} and that of the $D^*D\pi$ coupling is taken from~\cite{Baru:2013rta,Becirevic:2012pf}.
To reliably extract the $T$-matrix pole, the amplitude needs to be put into a finite volume to determine the parameters directly from a fit to the lattice energy levels; however, to demonstrate
the effect of the lhc it is sufficient to fit to the existing lattice data. The results are shown in 
Fig.~\ref{fig:fit} in red with orange uncertainty band~\footnote{These results were obtained with the $D$-meson recoil terms included, however their impact is negligible since the three-body cut (\ref{E3}) is sufficiently remote.}. 
Uncertainty in 
\cite{Padmanath:2022cvl} is given by the probability distribution for each phase shift data point. The nonzero probability of real phase shift data below the lhc is an artifact of the lhc being ignored in \cite{Padmanath:2022cvl}. 
Therefore, the lowest data point in \cite{Padmanath:2022cvl} is discarded in our analysis. We perform two fits: one where we use the part of the second data point 
that is above the lhc
(left panel of Fig.~\ref{fig:fit}) and the other one where we ignore it (right panel of Fig.~\ref{fig:fit}).
The upper three data points are included in both fits.

The poles of the $T$-matrix are now extracted using (\ref{poledef}).
From the left panel of Fig.~\ref{fig:fit} we conclude that, for the majority of the $1\sigma$ parameter space including the best fit, the amplitude contains two virtual states, both residing closer to the $DD^*$ threshold 
than the pole extracted in~\cite{Padmanath:2022cvl}, where only a single virtual state was found.
Those fits within the $1\sigma$ band, where $p \cot\delta$ does not cross the $ip$ curve above the lhc, describe the presence of a 
very narrow resonance showing up as a pair of complex poles below the $DD^*$
threshold. The appearance of a pair of virtual states is natural near the point where they are about to turn to a narrow resonance, as discussed in detail in \cite{McVoy:1968wht,Hanhart:2008mx,Guo:2009ct,Hanhart:2014ssa}. The position of the lhc sets the upper bound on the virtual pole binding energy---the collision between the virtual pole and its counterpart 
always takes place between the lhc branch point and the two-body threshold.
The fit results presented in the right panel of Fig.~\ref{fig:fit} are also consistent with the picture just drawn, with the narrow resonance scenario preferred. 

In both fits $p \cot\delta$ contains a near-threshold pole as a result of a subtle interplay of the repulsive OPE and the attractive short-range interaction. According to (\ref{phvsT}), this pole manifests itself as a zero in the $T$-matrix and provides yet another illustration that, in the current setting, $p \cot\delta$ cannot be approximated by a polynomial with a finite number of terms.
Moreover, it emphasizes the important role played by the OPE for understanding the analytic structure of the scattering amplitude.  
The existence of such a pole in $p\cot\delta$ for the neutron-deuteron scattering was already investigated half a century ago~\cite{Phillips:1969hm,Reiner:1969mmv}.
Here the location of the pole in $p\cot\delta$ very close to the lhc branch point is dictated by the particular lattice data points taken from \cite{Padmanath:2022cvl}. Indeed, 
while the lhc branch point depends only on the masses involved,
the exact location of the pole is sensitive to the interaction strength. If the pion coupling constant is artificially decreased or increased by some factor and the lattice data are refitted to fix the contact terms, the pole in $p\cot\delta$ either disappears below the lhc or moves closer to the two-body threshold~\cite{supp}. In the latter case, it may impose a stronger constraint on the ERE convergence radius than the lhc branch point. 

{\it Concluding remarks.}---An interesting question is what our findings imply for the nature
of the $\Tcc$ given an intriguing relationship between the location of
a near-threshold bound state pole and the size of the molecular
component \cite{Weinberg:1965zz,Morgan:1992ge}~\footnote{The equivalence of the two approaches was first discussed in \cite{Baru:2003qq}, 
generalization to virtual states is provided in 
\cite{Matuschek:2020gqe} and, for a positive effective range to be the case for the lattice analyses of $\Tcc$ \cite{Padmanath:2022cvl,Chen:2022vpo}, in \cite{Li:2021cue,Albaladejo:2022sux}.}.
This formalism relies on the
assumption that the binding momentum $\gamma$ is by far the smallest scale
in the problem. However, the lhc induces an additional small scale,
which not only strongly limits the range of applicability of ERE (see \cite{Braaten:2020nmc} for a related discussion), but also calls for an improvement of
the entire Weinberg approach to describe the compositeness of a hadronic state~\footnote{That zeros in the $T$-matrix invalidate the original 
Weinberg formalism was discussed in \cite{Weinberg:1965zz,Baru:2010ww,Kang:2016jxw}.}. On the other hand, a clear visibility in the amplitude of the lhc induced by the OPE provides an additional strong support for the molecular nature of the $T_{cc}(3875)^+$.

In summary, we have demonstrated that the lhcs from the long-range interactions, with the nearest one originating from the OPE, 
strongly restrict the range of validity of ERE in extracting the pole positions of near-threshold states. In order to increase this range and thereby extract the poles reliably, at least the OPE needs to be considered explicitly. Our findings clearly suggest that a direct comparison of the below-threshold energy levels predicted by our formalism, put in a finite volume, with the lattice energy levels should be made.
The described feature is general, and we illustrated it
by reanalyzing the lattice data of \cite{Padmanath:2022cvl} for the $T_{cc}(3875)^+$. The same effect may be operative in other hadronic systems where the lhc appears within the range covered by the theory and should be taken into account not only in lattice studies but also in analytical approaches already for physical parameters.
These might include not only $BB^*$ and $B\bar{B}^*$, but also $\Sigma(\Lambda) N$, $\Sigma_c^{(*)} \bar D^{(*)}$, $N D^{(*)}$, $N \bar D^{(*)}$ systems and so forth. Consequently, the effect of the lhc could be of relevance for probing the $\Sigma$ hyper-nuclei, $D$-mesic nuclei and the properties of charmed and bottom mesons in nuclear matter. Moreover, the $D\bar{D}^*$ is of particular interest for its obvious similarity to the $T_{cc}^+$ case studied above. The lattice simulations performed in~\cite{Prelovsek:2013cra} predict the $X(3872)$ as a shallow bound state $(11\pm 7)$~MeV below the $D\bar{D}^*$ threshold. Within the uncertainty, the $X$ pole may appear either below or above the lhc, so the influence of the latter is hard to foresee, and a reanalysis of the system on the lattice seems necessary. 

Lattice QCD enters the era of precise calculations for low-lying near-threshold states, which have been a focus of experimental and theoretical hadron physics for two decades and will clearly remain a hot topic in the foreseeable future. Our finding can be crucial in accurately extracting the pole locations of the benchmark systems such as $DD^*$ and $D\bar D^*$ from lattice data. Thus, the consequences of taking it into account are potentially very important for understanding various exotic hadrons, {\it e.g.}, whether they exist and what their internal structure is, from first principles of QCD.


\begin{acknowledgments}

The authors would like to thank Sasa Prelovsek for reading the manuscript and valuable comments.
This work is supported in part by the National Natural Science Foundation of China (NSFC) and the Deutsche Forschungsgemeinschaft (DFG) through the funds provided to the Sino-German Collaborative Research Center TRR110 ``Symmetries and the Emergence of Structure in QCD'' (NSFC Grant No. 12070131001, DFG Project-ID 196253076); by the Chinese Academy of Sciences under Grant No. YSBR-101 and No. XDB34030000; by the NSFC under Grants No. 12125507, No. 11835015, and No. 12047503; by the EU STRONG-2020 project under the program
H2020-INFRAIA-2018-1 with Grant No. 824093; by the Spanish Ministerio de Ciencia e Innovación (MICINN) under Grant No. PID2020-112777GB-I00; by Generalitat Valenciana under Grant PROMETEO/2020/023; by BMBF (Contract No. 05P21PCFP1) and by the MKW NRW under the funding code NW21-024-A. A.N. is supported by the Slovenian
Research Agency (research core Funding No. P1-0035). Q.W. is also supported by Guangdong Provincial funding with Grant No.~2019QN01X172.

\end{acknowledgments}

\bibliography{Tcc_refs}

\begin{thebibliography}{60}%
\makeatletter
\providecommand \@ifxundefined [1]{%
 \@ifx{#1\undefined}
}%
\providecommand \@ifnum [1]{%
 \ifnum #1\expandafter \@firstoftwo
 \else \expandafter \@secondoftwo
 \fi
}%
\providecommand \@ifx [1]{%
 \ifx #1\expandafter \@firstoftwo
 \else \expandafter \@secondoftwo
 \fi
}%
\providecommand \natexlab [1]{#1}%
\providecommand \enquote  [1]{``#1''}%
\providecommand \bibnamefont  [1]{#1}%
\providecommand \bibfnamefont [1]{#1}%
\providecommand \citenamefont [1]{#1}%
\providecommand \href@noop [0]{\@secondoftwo}%
\providecommand \href [0]{\begingroup \@sanitize@url \@href}%
\providecommand \@href[1]{\@@startlink{#1}\@@href}%
\providecommand \@@href[1]{\endgroup#1\@@endlink}%
\providecommand \@sanitize@url [0]{\catcode `\\12\catcode `\$12\catcode
  `\&12\catcode `\#12\catcode `\^12\catcode `\_12\catcode `\%12\relax}%
\providecommand \@@startlink[1]{}%
\providecommand \@@endlink[0]{}%
\providecommand \url  [0]{\begingroup\@sanitize@url \@url }%
\providecommand \@url [1]{\endgroup\@href {#1}{\urlprefix }}%
\providecommand \urlprefix  [0]{URL }%
\providecommand \Eprint [0]{\href }%
\providecommand \doibase [0]{https://doi.org/}%
\providecommand \selectlanguage [0]{\@gobble}%
\providecommand \bibinfo  [0]{\@secondoftwo}%
\providecommand \bibfield  [0]{\@secondoftwo}%
\providecommand \translation [1]{[#1]}%
\providecommand \BibitemOpen [0]{}%
\providecommand \bibitemStop [0]{}%
\providecommand \bibitemNoStop [0]{.\EOS\space}%
\providecommand \EOS [0]{\spacefactor3000\relax}%
\providecommand \BibitemShut  [1]{\csname bibitem#1\endcsname}%
\let\auto@bib@innerbib\@empty
\bibitem [{\citenamefont {Esposito}\ \emph {et~al.}(2015)\citenamefont
  {Esposito}, \citenamefont {Guerrieri}, \citenamefont {Piccinini},
  \citenamefont {Pilloni},\ and\ \citenamefont {Polosa}}]{Esposito:2014rxa}%
  \BibitemOpen
  \bibfield  {author} {\bibinfo {author} {\bibfnamefont {A.}~\bibnamefont
  {Esposito}}, \bibinfo {author} {\bibfnamefont {A.~L.}\ \bibnamefont
  {Guerrieri}}, \bibinfo {author} {\bibfnamefont {F.}~\bibnamefont
  {Piccinini}}, \bibinfo {author} {\bibfnamefont {A.}~\bibnamefont {Pilloni}},\
  and\ \bibinfo {author} {\bibfnamefont {A.~D.}\ \bibnamefont {Polosa}},\
  }\bibfield  {title} {\bibinfo {title} {{Four-Quark Hadrons: an Updated
  Review}},\ }\href {https://doi.org/10.1142/S0217751X15300021} {\bibfield
  {journal} {\bibinfo  {journal} {Int. J. Mod. Phys. A}\ }\textbf {\bibinfo
  {volume} {30}},\ \bibinfo {pages} {1530002} (\bibinfo {year} {2015})},\
  \Eprint {https://arxiv.org/abs/1411.5997} {arXiv:1411.5997 [hep-ph]}
  \BibitemShut {NoStop}%
\bibitem [{\citenamefont {Lebed}\ \emph {et~al.}(2017)\citenamefont {Lebed},
  \citenamefont {Mitchell},\ and\ \citenamefont {Swanson}}]{Lebed:2016hpi}%
  \BibitemOpen
  \bibfield  {author} {\bibinfo {author} {\bibfnamefont {R.~F.}\ \bibnamefont
  {Lebed}}, \bibinfo {author} {\bibfnamefont {R.~E.}\ \bibnamefont
  {Mitchell}},\ and\ \bibinfo {author} {\bibfnamefont {E.~S.}\ \bibnamefont
  {Swanson}},\ }\bibfield  {title} {\bibinfo {title} {{Heavy-Quark QCD
  Exotica}},\ }\href {https://doi.org/10.1016/j.ppnp.2016.11.003} {\bibfield
  {journal} {\bibinfo  {journal} {Prog. Part. Nucl. Phys.}\ }\textbf {\bibinfo
  {volume} {93}},\ \bibinfo {pages} {143} (\bibinfo {year} {2017})},\ \Eprint
  {https://arxiv.org/abs/1610.04528} {arXiv:1610.04528 [hep-ph]} \BibitemShut
  {NoStop}%
\bibitem [{\citenamefont {Chen}\ \emph {et~al.}(2016)\citenamefont {Chen},
  \citenamefont {Chen}, \citenamefont {Liu},\ and\ \citenamefont
  {Zhu}}]{Chen:2016qju}%
  \BibitemOpen
  \bibfield  {author} {\bibinfo {author} {\bibfnamefont {H.-X.}\ \bibnamefont
  {Chen}}, \bibinfo {author} {\bibfnamefont {W.}~\bibnamefont {Chen}}, \bibinfo
  {author} {\bibfnamefont {X.}~\bibnamefont {Liu}},\ and\ \bibinfo {author}
  {\bibfnamefont {S.-L.}\ \bibnamefont {Zhu}},\ }\bibfield  {title} {\bibinfo
  {title} {{The hidden-charm pentaquark and tetraquark states}},\ }\href
  {https://doi.org/10.1016/j.physrep.2016.05.004} {\bibfield  {journal}
  {\bibinfo  {journal} {Phys. Rept.}\ }\textbf {\bibinfo {volume} {639}},\
  \bibinfo {pages} {1} (\bibinfo {year} {2016})},\ \Eprint
  {https://arxiv.org/abs/1601.02092} {arXiv:1601.02092 [hep-ph]} \BibitemShut
  {NoStop}%
\bibitem [{\citenamefont {Guo}\ \emph {et~al.}(2018)\citenamefont {Guo},
  \citenamefont {Hanhart}, \citenamefont {Mei\ss{}ner}, \citenamefont {Wang},
  \citenamefont {Zhao},\ and\ \citenamefont {Zou}}]{Guo:2017jvc}%
  \BibitemOpen
  \bibfield  {author} {\bibinfo {author} {\bibfnamefont {F.-K.}\ \bibnamefont
  {Guo}}, \bibinfo {author} {\bibfnamefont {C.}~\bibnamefont {Hanhart}},
  \bibinfo {author} {\bibfnamefont {U.-G.}\ \bibnamefont {Mei\ss{}ner}},
  \bibinfo {author} {\bibfnamefont {Q.}~\bibnamefont {Wang}}, \bibinfo {author}
  {\bibfnamefont {Q.}~\bibnamefont {Zhao}},\ and\ \bibinfo {author}
  {\bibfnamefont {B.-S.}\ \bibnamefont {Zou}},\ }\bibfield  {title} {\bibinfo
  {title} {{Hadronic molecules}},\ }\href
  {https://doi.org/10.1103/RevModPhys.90.015004} {\bibfield  {journal}
  {\bibinfo  {journal} {Rev. Mod. Phys.}\ }\textbf {\bibinfo {volume} {90}},\
  \bibinfo {pages} {015004} (\bibinfo {year} {2018})},\ \Eprint
  {https://arxiv.org/abs/1705.00141} {arXiv:1705.00141 [hep-ph]} \BibitemShut
  {NoStop}%
\bibitem [{\citenamefont {Kalashnikova}\ and\ \citenamefont
  {Nefediev}(2019)}]{Kalashnikova:2018vkv}%
  \BibitemOpen
  \bibfield  {author} {\bibinfo {author} {\bibfnamefont {Y.~S.}\ \bibnamefont
  {Kalashnikova}}\ and\ \bibinfo {author} {\bibfnamefont {A.~V.}\ \bibnamefont
  {Nefediev}},\ }\bibfield  {title} {\bibinfo {title} {{X(3872) in the
  molecular model}},\ }\href {https://doi.org/10.3367/UFNe.2018.08.038411}
  {\bibfield  {journal} {\bibinfo  {journal} {Phys. Usp.}\ }\textbf {\bibinfo
  {volume} {62}},\ \bibinfo {pages} {568} (\bibinfo {year} {2019})},\ \Eprint
  {https://arxiv.org/abs/1811.01324} {arXiv:1811.01324 [hep-ph]} \BibitemShut
  {NoStop}%
\bibitem [{\citenamefont {Yamaguchi}\ \emph {et~al.}(2020)\citenamefont
  {Yamaguchi}, \citenamefont {Hosaka}, \citenamefont {Takeuchi},\ and\
  \citenamefont {Takizawa}}]{Yamaguchi:2019vea}%
  \BibitemOpen
  \bibfield  {author} {\bibinfo {author} {\bibfnamefont {Y.}~\bibnamefont
  {Yamaguchi}}, \bibinfo {author} {\bibfnamefont {A.}~\bibnamefont {Hosaka}},
  \bibinfo {author} {\bibfnamefont {S.}~\bibnamefont {Takeuchi}},\ and\
  \bibinfo {author} {\bibfnamefont {M.}~\bibnamefont {Takizawa}},\ }\bibfield
  {title} {\bibinfo {title} {{Heavy hadronic molecules with pion exchange and
  quark core couplings: a guide for practitioners}},\ }\href
  {https://doi.org/10.1088/1361-6471/ab72b0} {\bibfield  {journal} {\bibinfo
  {journal} {J. Phys. G}\ }\textbf {\bibinfo {volume} {47}},\ \bibinfo {pages}
  {053001} (\bibinfo {year} {2020})},\ \Eprint
  {https://arxiv.org/abs/1908.08790} {arXiv:1908.08790 [hep-ph]} \BibitemShut
  {NoStop}%
\bibitem [{\citenamefont {Brambilla}\ \emph {et~al.}(2020)\citenamefont
  {Brambilla}, \citenamefont {Eidelman}, \citenamefont {Hanhart}, \citenamefont
  {Nefediev}, \citenamefont {Shen}, \citenamefont {Thomas}, \citenamefont
  {Vairo},\ and\ \citenamefont {Yuan}}]{Brambilla:2019esw}%
  \BibitemOpen
  \bibfield  {author} {\bibinfo {author} {\bibfnamefont {N.}~\bibnamefont
  {Brambilla}}, \bibinfo {author} {\bibfnamefont {S.}~\bibnamefont {Eidelman}},
  \bibinfo {author} {\bibfnamefont {C.}~\bibnamefont {Hanhart}}, \bibinfo
  {author} {\bibfnamefont {A.}~\bibnamefont {Nefediev}}, \bibinfo {author}
  {\bibfnamefont {C.-P.}\ \bibnamefont {Shen}}, \bibinfo {author}
  {\bibfnamefont {C.~E.}\ \bibnamefont {Thomas}}, \bibinfo {author}
  {\bibfnamefont {A.}~\bibnamefont {Vairo}},\ and\ \bibinfo {author}
  {\bibfnamefont {C.-Z.}\ \bibnamefont {Yuan}},\ }\bibfield  {title} {\bibinfo
  {title} {{The $XYZ$ states: experimental and theoretical status and
  perspectives}},\ }\href {https://doi.org/10.1016/j.physrep.2020.05.001}
  {\bibfield  {journal} {\bibinfo  {journal} {Phys. Rept.}\ }\textbf {\bibinfo
  {volume} {873}},\ \bibinfo {pages} {1} (\bibinfo {year} {2020})},\ \Eprint
  {https://arxiv.org/abs/1907.07583} {arXiv:1907.07583 [hep-ex]} \BibitemShut
  {NoStop}%
\bibitem [{\citenamefont {Guo}\ \emph {et~al.}(2020)\citenamefont {Guo},
  \citenamefont {Liu},\ and\ \citenamefont {Sakai}}]{Guo:2019twa}%
  \BibitemOpen
  \bibfield  {author} {\bibinfo {author} {\bibfnamefont {F.-K.}\ \bibnamefont
  {Guo}}, \bibinfo {author} {\bibfnamefont {X.-H.}\ \bibnamefont {Liu}},\ and\
  \bibinfo {author} {\bibfnamefont {S.}~\bibnamefont {Sakai}},\ }\bibfield
  {title} {\bibinfo {title} {{Threshold cusps and triangle singularities in
  hadronic reactions}},\ }\href {https://doi.org/10.1016/j.ppnp.2020.103757}
  {\bibfield  {journal} {\bibinfo  {journal} {Prog. Part. Nucl. Phys.}\
  }\textbf {\bibinfo {volume} {112}},\ \bibinfo {pages} {103757} (\bibinfo
  {year} {2020})},\ \Eprint {https://arxiv.org/abs/1912.07030}
  {arXiv:1912.07030 [hep-ph]} \BibitemShut {NoStop}%
\bibitem [{\citenamefont {Chen}\ \emph {et~al.}(2023)\citenamefont {Chen},
  \citenamefont {Chen}, \citenamefont {Liu}, \citenamefont {Liu},\ and\
  \citenamefont {Zhu}}]{Chen:2022asf}%
  \BibitemOpen
  \bibfield  {author} {\bibinfo {author} {\bibfnamefont {H.-X.}\ \bibnamefont
  {Chen}}, \bibinfo {author} {\bibfnamefont {W.}~\bibnamefont {Chen}}, \bibinfo
  {author} {\bibfnamefont {X.}~\bibnamefont {Liu}}, \bibinfo {author}
  {\bibfnamefont {Y.-R.}\ \bibnamefont {Liu}},\ and\ \bibinfo {author}
  {\bibfnamefont {S.-L.}\ \bibnamefont {Zhu}},\ }\bibfield  {title} {\bibinfo
  {title} {{An updated review of the new hadron states}},\ }\href
  {https://doi.org/10.1088/1361-6633/aca3b6} {\bibfield  {journal} {\bibinfo
  {journal} {Rept. Prog. Phys.}\ }\textbf {\bibinfo {volume} {86}},\ \bibinfo
  {pages} {026201} (\bibinfo {year} {2023})},\ \Eprint
  {https://arxiv.org/abs/2204.02649} {arXiv:2204.02649 [hep-ph]} \BibitemShut
  {NoStop}%
\bibitem [{\citenamefont {Braaten}\ and\ \citenamefont
  {Lu}(2007)}]{Braaten:2007dw}%
  \BibitemOpen
  \bibfield  {author} {\bibinfo {author} {\bibfnamefont {E.}~\bibnamefont
  {Braaten}}\ and\ \bibinfo {author} {\bibfnamefont {M.}~\bibnamefont {Lu}},\
  }\bibfield  {title} {\bibinfo {title} {{Line shapes of the $X(3872)$}},\
  }\href {https://doi.org/10.1103/PhysRevD.76.094028} {\bibfield  {journal}
  {\bibinfo  {journal} {Phys. Rev. D}\ }\textbf {\bibinfo {volume} {76}},\
  \bibinfo {pages} {094028} (\bibinfo {year} {2007})},\ \Eprint
  {https://arxiv.org/abs/0709.2697} {arXiv:0709.2697 [hep-ph]} \BibitemShut
  {NoStop}%
\bibitem [{\citenamefont {Hanhart}\ \emph {et~al.}(2010)\citenamefont
  {Hanhart}, \citenamefont {Kalashnikova},\ and\ \citenamefont
  {Nefediev}}]{Hanhart:2010wh}%
  \BibitemOpen
  \bibfield  {author} {\bibinfo {author} {\bibfnamefont {C.}~\bibnamefont
  {Hanhart}}, \bibinfo {author} {\bibfnamefont {Y.~S.}\ \bibnamefont
  {Kalashnikova}},\ and\ \bibinfo {author} {\bibfnamefont {A.~V.}\ \bibnamefont
  {Nefediev}},\ }\bibfield  {title} {\bibinfo {title} {{Lineshapes for
  composite particles with unstable constituents}},\ }\href
  {https://doi.org/10.1103/PhysRevD.81.094028} {\bibfield  {journal} {\bibinfo
  {journal} {Phys. Rev. D}\ }\textbf {\bibinfo {volume} {81}},\ \bibinfo
  {pages} {094028} (\bibinfo {year} {2010})},\ \Eprint
  {https://arxiv.org/abs/1002.4097} {arXiv:1002.4097 [hep-ph]} \BibitemShut
  {NoStop}%
\bibitem [{\citenamefont {Filin}\ \emph {et~al.}(2010)\citenamefont {Filin},
  \citenamefont {Romanov}, \citenamefont {Baru}, \citenamefont {Hanhart},
  \citenamefont {Kalashnikova}, \citenamefont {Kudryavtsev}, \citenamefont
  {Mei{\ss}ner},\ and\ \citenamefont {Nefediev}}]{Filin:2010se}%
  \BibitemOpen
  \bibfield  {author} {\bibinfo {author} {\bibfnamefont {A.~A.}\ \bibnamefont
  {Filin}}, \bibinfo {author} {\bibfnamefont {A.}~\bibnamefont {Romanov}},
  \bibinfo {author} {\bibfnamefont {V.}~\bibnamefont {Baru}}, \bibinfo {author}
  {\bibfnamefont {C.}~\bibnamefont {Hanhart}}, \bibinfo {author} {\bibfnamefont
  {Y.~S.}\ \bibnamefont {Kalashnikova}}, \bibinfo {author} {\bibfnamefont
  {A.~E.}\ \bibnamefont {Kudryavtsev}}, \bibinfo {author} {\bibfnamefont
  {U.-G.}\ \bibnamefont {Mei{\ss}ner}},\ and\ \bibinfo {author} {\bibfnamefont
  {A.~V.}\ \bibnamefont {Nefediev}},\ }\bibfield  {title} {\bibinfo {title}
  {{Comment on ``Possibility of Deeply Bound Hadronic Molecules from Single
  Pion Exchange''}},\ }\href {https://doi.org/10.1103/PhysRevLett.105.019101}
  {\bibfield  {journal} {\bibinfo  {journal} {Phys. Rev. Lett.}\ }\textbf
  {\bibinfo {volume} {105}},\ \bibinfo {pages} {019101} (\bibinfo {year}
  {2010})},\ \Eprint {https://arxiv.org/abs/1004.4789} {arXiv:1004.4789
  [hep-ph]} \BibitemShut {NoStop}%
\bibitem [{\citenamefont {Guo}\ and\ \citenamefont
  {Mei{\ss}ner}(2011)}]{Guo:2011dd}%
  \BibitemOpen
  \bibfield  {author} {\bibinfo {author} {\bibfnamefont {F.-K.}\ \bibnamefont
  {Guo}}\ and\ \bibinfo {author} {\bibfnamefont {U.-G.}\ \bibnamefont
  {Mei{\ss}ner}},\ }\bibfield  {title} {\bibinfo {title} {{More kaonic bound
  states and a comprehensive interpretation of the $D_{sJ}$ states}},\ }\href
  {https://doi.org/10.1103/PhysRevD.84.014013} {\bibfield  {journal} {\bibinfo
  {journal} {Phys. Rev. D}\ }\textbf {\bibinfo {volume} {84}},\ \bibinfo
  {pages} {014013} (\bibinfo {year} {2011})},\ \Eprint
  {https://arxiv.org/abs/1102.3536} {arXiv:1102.3536 [hep-ph]} \BibitemShut
  {NoStop}%
\bibitem [{\citenamefont {Aaij}\ \emph
  {et~al.}(2022{\natexlab{a}})\citenamefont {Aaij} \emph
  {et~al.}}]{LHCb:2021vvq}%
  \BibitemOpen
  \bibfield  {author} {\bibinfo {author} {\bibfnamefont {R.}~\bibnamefont
  {Aaij}} \emph {et~al.} (\bibinfo {collaboration} {LHCb}),\ }\bibfield
  {title} {\bibinfo {title} {{Observation of an exotic narrow doubly charmed
  tetraquark}},\ }\href {https://doi.org/10.1038/s41567-022-01614-y} {\bibfield
   {journal} {\bibinfo  {journal} {Nature Phys.}\ }\textbf {\bibinfo {volume}
  {18}},\ \bibinfo {pages} {751} (\bibinfo {year} {2022}{\natexlab{a}})},\
  \Eprint {https://arxiv.org/abs/2109.01038} {arXiv:2109.01038 [hep-ex]}
  \BibitemShut {NoStop}%
\bibitem [{\citenamefont {Aaij}\ \emph
  {et~al.}(2022{\natexlab{b}})\citenamefont {Aaij} \emph
  {et~al.}}]{LHCb:2021auc}%
  \BibitemOpen
  \bibfield  {author} {\bibinfo {author} {\bibfnamefont {R.}~\bibnamefont
  {Aaij}} \emph {et~al.} (\bibinfo {collaboration} {LHCb}),\ }\bibfield
  {title} {\bibinfo {title} {{Study of the doubly charmed tetraquark
  $T_{cc}^{+}$}},\ }\href {https://doi.org/10.1038/s41467-022-30206-w}
  {\bibfield  {journal} {\bibinfo  {journal} {Nature Commun.}\ }\textbf
  {\bibinfo {volume} {13}},\ \bibinfo {pages} {3351} (\bibinfo {year}
  {2022}{\natexlab{b}})},\ \Eprint {https://arxiv.org/abs/2109.01056}
  {arXiv:2109.01056 [hep-ex]} \BibitemShut {NoStop}%
\bibitem [{\citenamefont {Blankenbecler}\ \emph {et~al.}(1961)\citenamefont
  {Blankenbecler}, \citenamefont {Goldberger}, \citenamefont {MacDowell},\ and\
  \citenamefont {Treiman}}]{BGMT}%
  \BibitemOpen
  \bibfield  {author} {\bibinfo {author} {\bibfnamefont {R.}~\bibnamefont
  {Blankenbecler}}, \bibinfo {author} {\bibfnamefont {M.~L.}\ \bibnamefont
  {Goldberger}}, \bibinfo {author} {\bibfnamefont {S.~W.}\ \bibnamefont
  {MacDowell}},\ and\ \bibinfo {author} {\bibfnamefont {S.~B.}\ \bibnamefont
  {Treiman}},\ }\bibfield  {title} {\bibinfo {title} {{Singularities of
  Scattering Amplitudes on Unphysical Sheets and Their Interpretation}},\
  }\href {https://doi.org/10.1103/PhysRev.123.692} {\bibfield  {journal}
  {\bibinfo  {journal} {Phys. Rev.}\ }\textbf {\bibinfo {volume} {123}},\
  \bibinfo {pages} {692} (\bibinfo {year} {1961})}\BibitemShut {NoStop}%
\bibitem [{\citenamefont {D{\"o}ring}\ \emph {et~al.}(2009)\citenamefont
  {D{\"o}ring}, \citenamefont {Hanhart}, \citenamefont {Huang}, \citenamefont
  {Krewald},\ and\ \citenamefont {Mei{\ss}ner}}]{Doring:2009yv}%
  \BibitemOpen
  \bibfield  {author} {\bibinfo {author} {\bibfnamefont {M.}~\bibnamefont
  {D{\"o}ring}}, \bibinfo {author} {\bibfnamefont {C.}~\bibnamefont {Hanhart}},
  \bibinfo {author} {\bibfnamefont {F.}~\bibnamefont {Huang}}, \bibinfo
  {author} {\bibfnamefont {S.}~\bibnamefont {Krewald}},\ and\ \bibinfo {author}
  {\bibfnamefont {U.-G.}\ \bibnamefont {Mei{\ss}ner}},\ }\bibfield  {title}
  {\bibinfo {title} {{Analytic properties of the scattering amplitude and
  resonances parameters in a meson exchange model}},\ }\href
  {https://doi.org/10.1016/j.nuclphysa.2009.08.010} {\bibfield  {journal}
  {\bibinfo  {journal} {Nucl. Phys. A}\ }\textbf {\bibinfo {volume} {829}},\
  \bibinfo {pages} {170} (\bibinfo {year} {2009})},\ \Eprint
  {https://arxiv.org/abs/0903.4337} {arXiv:0903.4337 [nucl-th]} \BibitemShut
  {NoStop}%
\bibitem [{\citenamefont {Frautschi}\ and\ \citenamefont
  {Walecka}(1960)}]{Frautschi:1960qzm}%
  \BibitemOpen
  \bibfield  {author} {\bibinfo {author} {\bibfnamefont {S.~C.}\ \bibnamefont
  {Frautschi}}\ and\ \bibinfo {author} {\bibfnamefont {J.~D.}\ \bibnamefont
  {Walecka}},\ }\bibfield  {title} {\bibinfo {title} {{Pion-Nucleon Scattering
  in the Mandelstam Representation}},\ }\href
  {https://doi.org/10.1103/PhysRev.120.1486} {\bibfield  {journal} {\bibinfo
  {journal} {Phys. Rev.}\ }\textbf {\bibinfo {volume} {120}},\ \bibinfo {pages}
  {1486} (\bibinfo {year} {1960})}\BibitemShut {NoStop}%
\bibitem [{\citenamefont {Oller}(2019)}]{Oller:2019rej}%
  \BibitemOpen
  \bibfield  {author} {\bibinfo {author} {\bibfnamefont {J.~A.}\ \bibnamefont
  {Oller}},\ }\href {https://doi.org/10.1007/978-3-030-13582-9} {\emph
  {\bibinfo {title} {{A Brief Introduction to Dispersion Relations}}}},\
  SpringerBriefs in Physics\ (\bibinfo  {publisher} {Springer},\ \bibinfo
  {year} {2019})\BibitemShut {NoStop}%
\bibitem [{\citenamefont {Omn\`es}(1971)}]{omnes1971}%
  \BibitemOpen
  \bibfield  {author} {\bibinfo {author} {\bibfnamefont {R.}~\bibnamefont
  {Omn\`es}},\ }\href@noop {} {\emph {\bibinfo {title} {Introduction to
  Particle Physics}}}\ (\bibinfo  {publisher} {John Wiley \& Sons Ltd},\
  \bibinfo {address} {London},\ \bibinfo {year} {1971})\BibitemShut {NoStop}%
\bibitem [{\citenamefont {Padmanath}\ and\ \citenamefont
  {Prelovsek}(2022)}]{Padmanath:2022cvl}%
  \BibitemOpen
  \bibfield  {author} {\bibinfo {author} {\bibfnamefont {M.}~\bibnamefont
  {Padmanath}}\ and\ \bibinfo {author} {\bibfnamefont {S.}~\bibnamefont
  {Prelovsek}},\ }\bibfield  {title} {\bibinfo {title} {{Signature of a Doubly
  Charm Tetraquark Pole in $DD^*$ Scattering on the Lattice}},\ }\href
  {https://doi.org/10.1103/PhysRevLett.129.032002} {\bibfield  {journal}
  {\bibinfo  {journal} {Phys. Rev. Lett.}\ }\textbf {\bibinfo {volume} {129}},\
  \bibinfo {pages} {032002} (\bibinfo {year} {2022})},\ \Eprint
  {https://arxiv.org/abs/2202.10110} {arXiv:2202.10110 [hep-lat]} \BibitemShut
  {NoStop}%
\bibitem [{\citenamefont {Chen}\ \emph {et~al.}(2022)\citenamefont {Chen},
  \citenamefont {Shi}, \citenamefont {Chen}, \citenamefont {Gong},
  \citenamefont {Liu}, \citenamefont {Sun},\ and\ \citenamefont
  {Zhang}}]{Chen:2022vpo}%
  \BibitemOpen
  \bibfield  {author} {\bibinfo {author} {\bibfnamefont {S.}~\bibnamefont
  {Chen}}, \bibinfo {author} {\bibfnamefont {C.}~\bibnamefont {Shi}}, \bibinfo
  {author} {\bibfnamefont {Y.}~\bibnamefont {Chen}}, \bibinfo {author}
  {\bibfnamefont {M.}~\bibnamefont {Gong}}, \bibinfo {author} {\bibfnamefont
  {Z.}~\bibnamefont {Liu}}, \bibinfo {author} {\bibfnamefont {W.}~\bibnamefont
  {Sun}},\ and\ \bibinfo {author} {\bibfnamefont {R.}~\bibnamefont {Zhang}},\
  }\bibfield  {title} {\bibinfo {title} {{$T_{cc}^+(3875)$ relevant $DD^*$
  scattering from $N_f=2$ lattice QCD}},\ }\href
  {https://doi.org/10.1016/j.physletb.2022.137391} {\bibfield  {journal}
  {\bibinfo  {journal} {Phys. Lett. B}\ }\textbf {\bibinfo {volume} {833}},\
  \bibinfo {pages} {137391} (\bibinfo {year} {2022})},\ \Eprint
  {https://arxiv.org/abs/2206.06185} {arXiv:2206.06185 [hep-lat]} \BibitemShut
  {NoStop}%
\bibitem [{\citenamefont {Lyu}\ \emph {et~al.}(2023)\citenamefont {Lyu},
  \citenamefont {Aoki}, \citenamefont {Doi}, \citenamefont {Hatsuda},
  \citenamefont {Ikeda},\ and\ \citenamefont {Meng}}]{Lyu:2023xro}%
  \BibitemOpen
  \bibfield  {author} {\bibinfo {author} {\bibfnamefont {Y.}~\bibnamefont
  {Lyu}}, \bibinfo {author} {\bibfnamefont {S.}~\bibnamefont {Aoki}}, \bibinfo
  {author} {\bibfnamefont {T.}~\bibnamefont {Doi}}, \bibinfo {author}
  {\bibfnamefont {T.}~\bibnamefont {Hatsuda}}, \bibinfo {author} {\bibfnamefont
  {Y.}~\bibnamefont {Ikeda}},\ and\ \bibinfo {author} {\bibfnamefont
  {J.}~\bibnamefont {Meng}},\ }\bibfield  {title} {\bibinfo {title} {{Doubly
  charmed tetraquark $T^+_{cc}$ from Lattice QCD near Physical Point}},\
  }\href@noop {} {\  (\bibinfo {year} {2023})},\ \Eprint
  {https://arxiv.org/abs/2302.04505} {arXiv:2302.04505 [hep-lat]} \BibitemShut
  {NoStop}%
\bibitem [{Note1()}]{Note1}%
  \BibitemOpen
  \bibinfo {note} {If one of the scattering particles is unstable, it is
  convenient to make ERE around the complex branch point connected to the
  two-body channels~\cite {Braaten:2009jke,Baru:2021ldu}.}\BibitemShut {Stop}%
\bibitem [{\citenamefont {Raposo}\ and\ \citenamefont
  {Hansen}(2023)}]{Raposo:2023nex}%
  \BibitemOpen
  \bibfield  {author} {\bibinfo {author} {\bibfnamefont {A.~B.}\ \bibnamefont
  {Raposo}}\ and\ \bibinfo {author} {\bibfnamefont {M.~T.}\ \bibnamefont
  {Hansen}},\ }\bibfield  {title} {\bibinfo {title} {{The L\"uscher scattering
  formalism on the $t$-channel cut}},\ }\href
  {https://doi.org/10.22323/1.430.0051} {\bibfield  {journal} {\bibinfo
  {journal} {PoS}\ }\textbf {\bibinfo {volume} {LATTICE2022}},\ \bibinfo
  {pages} {051} (\bibinfo {year} {2023})},\ \Eprint
  {https://arxiv.org/abs/2301.03981} {arXiv:2301.03981 [hep-lat]} \BibitemShut
  {NoStop}%
\bibitem [{\citenamefont {Dawid}\ \emph {et~al.}(2023)\citenamefont {Dawid},
  \citenamefont {Islam},\ and\ \citenamefont {Brice\~no}}]{Dawid:2023jrj}%
  \BibitemOpen
  \bibfield  {author} {\bibinfo {author} {\bibfnamefont {S.~M.}\ \bibnamefont
  {Dawid}}, \bibinfo {author} {\bibfnamefont {M.~H.~E.}\ \bibnamefont
  {Islam}},\ and\ \bibinfo {author} {\bibfnamefont {R.~A.}\ \bibnamefont
  {Brice\~no}},\ }\bibfield  {title} {\bibinfo {title} {{Analytic continuation
  of the relativistic three-particle scattering amplitudes}},\ }\href@noop {}
  {\  (\bibinfo {year} {2023})},\ \Eprint {https://arxiv.org/abs/2303.04394}
  {arXiv:2303.04394 [nucl-th]} \BibitemShut {NoStop}%
\bibitem [{\citenamefont {Meng}\ and\ \citenamefont
  {Epelbaum}(2021)}]{Meng:2021uhz}%
  \BibitemOpen
  \bibfield  {author} {\bibinfo {author} {\bibfnamefont {L.}~\bibnamefont
  {Meng}}\ and\ \bibinfo {author} {\bibfnamefont {E.}~\bibnamefont
  {Epelbaum}},\ }\bibfield  {title} {\bibinfo {title} {{Two-particle scattering
  from finite-volume quantization conditions using the plane wave basis}},\
  }\href {https://doi.org/10.1007/JHEP10(2021)051} {\bibfield  {journal}
  {\bibinfo  {journal} {JHEP}\ }\textbf {\bibinfo {volume} {10}},\ \bibinfo
  {pages} {051}},\ \Eprint {https://arxiv.org/abs/2108.02709} {arXiv:2108.02709
  [hep-lat]} \BibitemShut {NoStop}%
\bibitem [{\citenamefont {Baru}\ \emph {et~al.}(2011)\citenamefont {Baru},
  \citenamefont {Filin}, \citenamefont {Hanhart}, \citenamefont {Kalashnikova},
  \citenamefont {Kudryavtsev},\ and\ \citenamefont {Nefediev}}]{Baru:2011rs}%
  \BibitemOpen
  \bibfield  {author} {\bibinfo {author} {\bibfnamefont {V.}~\bibnamefont
  {Baru}}, \bibinfo {author} {\bibfnamefont {A.~A.}\ \bibnamefont {Filin}},
  \bibinfo {author} {\bibfnamefont {C.}~\bibnamefont {Hanhart}}, \bibinfo
  {author} {\bibfnamefont {Y.~S.}\ \bibnamefont {Kalashnikova}}, \bibinfo
  {author} {\bibfnamefont {A.~E.}\ \bibnamefont {Kudryavtsev}},\ and\ \bibinfo
  {author} {\bibfnamefont {A.~V.}\ \bibnamefont {Nefediev}},\ }\bibfield
  {title} {\bibinfo {title} {{Three-body $D\bar{D}\pi$ dynamics for the
  $X(3872)$}},\ }\href {https://doi.org/10.1103/PhysRevD.84.074029} {\bibfield
  {journal} {\bibinfo  {journal} {Phys. Rev. D}\ }\textbf {\bibinfo {volume}
  {84}},\ \bibinfo {pages} {074029} (\bibinfo {year} {2011})},\ \Eprint
  {https://arxiv.org/abs/1108.5644} {arXiv:1108.5644 [hep-ph]} \BibitemShut
  {NoStop}%
\bibitem [{\citenamefont {Schmidt}\ \emph {et~al.}(2018)\citenamefont
  {Schmidt}, \citenamefont {Jansen},\ and\ \citenamefont
  {Hammer}}]{Schmidt:2018vvl}%
  \BibitemOpen
  \bibfield  {author} {\bibinfo {author} {\bibfnamefont {M.}~\bibnamefont
  {Schmidt}}, \bibinfo {author} {\bibfnamefont {M.}~\bibnamefont {Jansen}},\
  and\ \bibinfo {author} {\bibfnamefont {H.-W.}\ \bibnamefont {Hammer}},\
  }\bibfield  {title} {\bibinfo {title} {{Threshold Effects and the Line Shape
  of the $X(3872)$ in Effective Field Theory}},\ }\href
  {https://doi.org/10.1103/PhysRevD.98.014032} {\bibfield  {journal} {\bibinfo
  {journal} {Phys. Rev. D}\ }\textbf {\bibinfo {volume} {98}},\ \bibinfo
  {pages} {014032} (\bibinfo {year} {2018})},\ \Eprint
  {https://arxiv.org/abs/1804.00375} {arXiv:1804.00375 [hep-ph]} \BibitemShut
  {NoStop}%
\bibitem [{\citenamefont {Braaten}\ \emph {et~al.}(2021)\citenamefont
  {Braaten}, \citenamefont {He},\ and\ \citenamefont
  {Jiang}}]{Braaten:2020nmc}%
  \BibitemOpen
  \bibfield  {author} {\bibinfo {author} {\bibfnamefont {E.}~\bibnamefont
  {Braaten}}, \bibinfo {author} {\bibfnamefont {L.-P.}\ \bibnamefont {He}},\
  and\ \bibinfo {author} {\bibfnamefont {J.}~\bibnamefont {Jiang}},\ }\bibfield
   {title} {\bibinfo {title} {{Galilean-invariant effective field theory for
  the $X(3872)$ at next-to-leading order}},\ }\href
  {https://doi.org/10.1103/PhysRevD.103.036014} {\bibfield  {journal} {\bibinfo
   {journal} {Phys. Rev. D}\ }\textbf {\bibinfo {volume} {103}},\ \bibinfo
  {pages} {036014} (\bibinfo {year} {2021})},\ \Eprint
  {https://arxiv.org/abs/2010.05801} {arXiv:2010.05801 [hep-ph]} \BibitemShut
  {NoStop}%
\bibitem [{\citenamefont {Du}\ \emph {et~al.}(2022)\citenamefont {Du},
  \citenamefont {Baru}, \citenamefont {Dong}, \citenamefont {Filin},
  \citenamefont {Guo}, \citenamefont {Hanhart}, \citenamefont {Nefediev},
  \citenamefont {Nieves},\ and\ \citenamefont {Wang}}]{Du:2021zzh}%
  \BibitemOpen
  \bibfield  {author} {\bibinfo {author} {\bibfnamefont {M.-L.}\ \bibnamefont
  {Du}}, \bibinfo {author} {\bibfnamefont {V.}~\bibnamefont {Baru}}, \bibinfo
  {author} {\bibfnamefont {X.-K.}\ \bibnamefont {Dong}}, \bibinfo {author}
  {\bibfnamefont {A.}~\bibnamefont {Filin}}, \bibinfo {author} {\bibfnamefont
  {F.-K.}\ \bibnamefont {Guo}}, \bibinfo {author} {\bibfnamefont
  {C.}~\bibnamefont {Hanhart}}, \bibinfo {author} {\bibfnamefont
  {A.}~\bibnamefont {Nefediev}}, \bibinfo {author} {\bibfnamefont
  {J.}~\bibnamefont {Nieves}},\ and\ \bibinfo {author} {\bibfnamefont
  {Q.}~\bibnamefont {Wang}},\ }\bibfield  {title} {\bibinfo {title}
  {{Coupled-channel approach to $T_{cc}^+$ including three-body effects}},\
  }\href {https://doi.org/10.1103/PhysRevD.105.014024} {\bibfield  {journal}
  {\bibinfo  {journal} {Phys. Rev. D}\ }\textbf {\bibinfo {volume} {105}},\
  \bibinfo {pages} {014024} (\bibinfo {year} {2022})},\ \Eprint
  {https://arxiv.org/abs/2110.13765} {arXiv:2110.13765 [hep-ph]} \BibitemShut
  {NoStop}%
\bibitem [{\citenamefont {Schweber}(1961)}]{Schweber:1961zz}%
  \BibitemOpen
  \bibfield  {author} {\bibinfo {author} {\bibfnamefont {S.~S.}\ \bibnamefont
  {Schweber}},\ }\href@noop {} {\emph {\bibinfo {title} {{An Introduction to
  Relativistic Quantum Field Theory}}}}\ (\bibinfo  {publisher} {Row, Peterson
  and Company},\ \bibinfo {address} {Evanston},\ \bibinfo {year}
  {1961})\BibitemShut {NoStop}%
\bibitem [{\citenamefont {Baru}\ \emph {et~al.}(2019)\citenamefont {Baru},
  \citenamefont {Epelbaum}, \citenamefont {Gegelia},\ and\ \citenamefont
  {Ren}}]{Baru:2019ndr}%
  \BibitemOpen
  \bibfield  {author} {\bibinfo {author} {\bibfnamefont {V.}~\bibnamefont
  {Baru}}, \bibinfo {author} {\bibfnamefont {E.}~\bibnamefont {Epelbaum}},
  \bibinfo {author} {\bibfnamefont {J.}~\bibnamefont {Gegelia}},\ and\ \bibinfo
  {author} {\bibfnamefont {X.~L.}\ \bibnamefont {Ren}},\ }\bibfield  {title}
  {\bibinfo {title} {{Towards baryon-baryon scattering in manifestly
  Lorentz-invariant formulation of SU(3) baryon chiral perturbation theory}},\
  }\href {https://doi.org/10.1016/j.physletb.2019.134987} {\bibfield  {journal}
  {\bibinfo  {journal} {Phys. Lett. B}\ }\textbf {\bibinfo {volume} {798}},\
  \bibinfo {pages} {134987} (\bibinfo {year} {2019})},\ \Eprint
  {https://arxiv.org/abs/1905.02116} {arXiv:1905.02116 [nucl-th]} \BibitemShut
  {NoStop}%
\bibitem [{sup()}]{supp}%
  \BibitemOpen
  \href@noop {} {}\bibinfo {note} {See Supplemental Material for technical
  details of the left-hand cuts, Lippmann-Schwinger equation, fitting procedure
  and additional analysis of unphysical cases.}\BibitemShut {Stop}%
\bibitem [{Note2()}]{Note2}%
  \BibitemOpen
  \bibinfo {note} {An important role played by the lhc from pion exchanges for
  $NN$ scattering at unphysical pion masses was pointed out in \cite
  {Baru:2015ira,Baru:2016evv}.}\BibitemShut {Stop}%
\bibitem [{\citenamefont {Gasser}\ and\ \citenamefont
  {Leutwyler}(1984)}]{Gasser:1983yg}%
  \BibitemOpen
  \bibfield  {author} {\bibinfo {author} {\bibfnamefont {J.}~\bibnamefont
  {Gasser}}\ and\ \bibinfo {author} {\bibfnamefont {H.}~\bibnamefont
  {Leutwyler}},\ }\bibfield  {title} {\bibinfo {title} {{Chiral Perturbation
  Theory to One Loop}},\ }\href {https://doi.org/10.1016/0003-4916(84)90242-2}
  {\bibfield  {journal} {\bibinfo  {journal} {Annals Phys.}\ }\textbf {\bibinfo
  {volume} {158}},\ \bibinfo {pages} {142} (\bibinfo {year}
  {1984})}\BibitemShut {NoStop}%
\bibitem [{\citenamefont {Baru}\ \emph {et~al.}(2013)\citenamefont {Baru},
  \citenamefont {Epelbaum}, \citenamefont {Filin}, \citenamefont {Hanhart},
  \citenamefont {Mei{\ss}ner},\ and\ \citenamefont {Nefediev}}]{Baru:2013rta}%
  \BibitemOpen
  \bibfield  {author} {\bibinfo {author} {\bibfnamefont {V.}~\bibnamefont
  {Baru}}, \bibinfo {author} {\bibfnamefont {E.}~\bibnamefont {Epelbaum}},
  \bibinfo {author} {\bibfnamefont {A.~A.}\ \bibnamefont {Filin}}, \bibinfo
  {author} {\bibfnamefont {C.}~\bibnamefont {Hanhart}}, \bibinfo {author}
  {\bibfnamefont {U.-G.}\ \bibnamefont {Mei{\ss}ner}},\ and\ \bibinfo {author}
  {\bibfnamefont {A.~V.}\ \bibnamefont {Nefediev}},\ }\bibfield  {title}
  {\bibinfo {title} {{Quark mass dependence of the $X(3872)$ binding energy}},\
  }\href {https://doi.org/10.1016/j.physletb.2013.08.073} {\bibfield  {journal}
  {\bibinfo  {journal} {Phys. Lett. B}\ }\textbf {\bibinfo {volume} {726}},\
  \bibinfo {pages} {537} (\bibinfo {year} {2013})},\ \Eprint
  {https://arxiv.org/abs/1306.4108} {arXiv:1306.4108 [hep-ph]} \BibitemShut
  {NoStop}%
\bibitem [{\citenamefont {Becirevic}\ and\ \citenamefont
  {Sanfilippo}(2013)}]{Becirevic:2012pf}%
  \BibitemOpen
  \bibfield  {author} {\bibinfo {author} {\bibfnamefont {D.}~\bibnamefont
  {Becirevic}}\ and\ \bibinfo {author} {\bibfnamefont {F.}~\bibnamefont
  {Sanfilippo}},\ }\bibfield  {title} {\bibinfo {title} {{Theoretical estimate
  of the $D^* \to D\pi$ decay rate}},\ }\href
  {https://doi.org/10.1016/j.physletb.2013.03.004} {\bibfield  {journal}
  {\bibinfo  {journal} {Phys. Lett. B}\ }\textbf {\bibinfo {volume} {721}},\
  \bibinfo {pages} {94} (\bibinfo {year} {2013})},\ \Eprint
  {https://arxiv.org/abs/1210.5410} {arXiv:1210.5410 [hep-lat]} \BibitemShut
  {NoStop}%
\bibitem [{Note3()}]{Note3}%
  \BibitemOpen
  \bibinfo {note} {These results were obtained with the $D$-meson recoil terms
  included, however their impact is negligible since the three-body cut (\ref
  {E3}) is sufficiently remote.}\BibitemShut {Stop}%
\bibitem [{\citenamefont {McVoy}(1968)}]{McVoy:1968wht}%
  \BibitemOpen
  \bibfield  {author} {\bibinfo {author} {\bibfnamefont {K.~W.}\ \bibnamefont
  {McVoy}},\ }\bibfield  {title} {\bibinfo {title} {{Virtual states and
  resonances}},\ }\href {https://doi.org/10.1016/0375-9474(68)90741-0}
  {\bibfield  {journal} {\bibinfo  {journal} {Nucl. Phys. A}\ }\textbf
  {\bibinfo {volume} {115}},\ \bibinfo {pages} {481} (\bibinfo {year}
  {1968})}\BibitemShut {NoStop}%
\bibitem [{\citenamefont {Hanhart}\ \emph {et~al.}(2008)\citenamefont
  {Hanhart}, \citenamefont {Pel{\'a}ez},\ and\ \citenamefont
  {R{\'i}os}}]{Hanhart:2008mx}%
  \BibitemOpen
  \bibfield  {author} {\bibinfo {author} {\bibfnamefont {C.}~\bibnamefont
  {Hanhart}}, \bibinfo {author} {\bibfnamefont {J.~R.}\ \bibnamefont
  {Pel{\'a}ez}},\ and\ \bibinfo {author} {\bibfnamefont {G.}~\bibnamefont
  {R{\'i}os}},\ }\bibfield  {title} {\bibinfo {title} {{Quark mass dependence
  of the rho and sigma from dispersion relations and Chiral Perturbation
  Theory}},\ }\href {https://doi.org/10.1103/PhysRevLett.100.152001} {\bibfield
   {journal} {\bibinfo  {journal} {Phys. Rev. Lett.}\ }\textbf {\bibinfo
  {volume} {100}},\ \bibinfo {pages} {152001} (\bibinfo {year} {2008})},\
  \Eprint {https://arxiv.org/abs/0801.2871} {arXiv:0801.2871 [hep-ph]}
  \BibitemShut {NoStop}%
\bibitem [{\citenamefont {Guo}\ \emph {et~al.}(2009)\citenamefont {Guo},
  \citenamefont {Hanhart},\ and\ \citenamefont {Mei{\ss}ner}}]{Guo:2009ct}%
  \BibitemOpen
  \bibfield  {author} {\bibinfo {author} {\bibfnamefont {F.-K.}\ \bibnamefont
  {Guo}}, \bibinfo {author} {\bibfnamefont {C.}~\bibnamefont {Hanhart}},\ and\
  \bibinfo {author} {\bibfnamefont {U.-G.}\ \bibnamefont {Mei{\ss}ner}},\
  }\bibfield  {title} {\bibinfo {title} {{Interactions between heavy mesons and
  Goldstone bosons from chiral dynamics}},\ }\href
  {https://doi.org/10.1140/epja/i2009-10762-1} {\bibfield  {journal} {\bibinfo
  {journal} {Eur. Phys. J. A}\ }\textbf {\bibinfo {volume} {40}},\ \bibinfo
  {pages} {171} (\bibinfo {year} {2009})},\ \Eprint
  {https://arxiv.org/abs/0901.1597} {arXiv:0901.1597 [hep-ph]} \BibitemShut
  {NoStop}%
\bibitem [{\citenamefont {Hanhart}\ \emph {et~al.}(2014)\citenamefont
  {Hanhart}, \citenamefont {Pel{\'a}ez},\ and\ \citenamefont
  {R\'ios}}]{Hanhart:2014ssa}%
  \BibitemOpen
  \bibfield  {author} {\bibinfo {author} {\bibfnamefont {C.}~\bibnamefont
  {Hanhart}}, \bibinfo {author} {\bibfnamefont {J.~R.}\ \bibnamefont
  {Pel{\'a}ez}},\ and\ \bibinfo {author} {\bibfnamefont {G.}~\bibnamefont
  {R\'ios}},\ }\bibfield  {title} {\bibinfo {title} {{Remarks on pole
  trajectories for resonances}},\ }\href
  {https://doi.org/10.1016/j.physletb.2014.11.011} {\bibfield  {journal}
  {\bibinfo  {journal} {Phys. Lett. B}\ }\textbf {\bibinfo {volume} {739}},\
  \bibinfo {pages} {375} (\bibinfo {year} {2014})},\ \Eprint
  {https://arxiv.org/abs/1407.7452} {arXiv:1407.7452 [hep-ph]} \BibitemShut
  {NoStop}%
\bibitem [{\citenamefont {Phillips}\ and\ \citenamefont
  {Barton}(1969)}]{Phillips:1969hm}%
  \BibitemOpen
  \bibfield  {author} {\bibinfo {author} {\bibfnamefont {A.~C.}\ \bibnamefont
  {Phillips}}\ and\ \bibinfo {author} {\bibfnamefont {G.}~\bibnamefont
  {Barton}},\ }\bibfield  {title} {\bibinfo {title} {{Relations between
  low-energy three nucleon observables}},\ }\href
  {https://doi.org/10.1016/0370-2693(69)90324-4} {\bibfield  {journal}
  {\bibinfo  {journal} {Phys. Lett. B}\ }\textbf {\bibinfo {volume} {28}},\
  \bibinfo {pages} {378} (\bibinfo {year} {1969})}\BibitemShut {NoStop}%
\bibitem [{\citenamefont {Reiner}(1969)}]{Reiner:1969mmv}%
  \BibitemOpen
  \bibfield  {author} {\bibinfo {author} {\bibfnamefont {A.~S.}\ \bibnamefont
  {Reiner}},\ }\bibfield  {title} {\bibinfo {title} {{On the anomalous
  effective range expansion for nucleon-deuteron scattering in the $S = 1/2$
  state}},\ }\href {https://doi.org/10.1016/0370-2693(69)90327-X} {\bibfield
  {journal} {\bibinfo  {journal} {Phys. Lett. B}\ }\textbf {\bibinfo {volume}
  {28}},\ \bibinfo {pages} {387} (\bibinfo {year} {1969})}\BibitemShut
  {NoStop}%
\bibitem [{\citenamefont {Weinberg}(1965)}]{Weinberg:1965zz}%
  \BibitemOpen
  \bibfield  {author} {\bibinfo {author} {\bibfnamefont {S.}~\bibnamefont
  {Weinberg}},\ }\bibfield  {title} {\bibinfo {title} {{Evidence That the
  Deuteron Is Not an Elementary Particle}},\ }\href
  {https://doi.org/10.1103/PhysRev.137.B672} {\bibfield  {journal} {\bibinfo
  {journal} {Phys. Rev.}\ }\textbf {\bibinfo {volume} {137}},\ \bibinfo {pages}
  {B672} (\bibinfo {year} {1965})}\BibitemShut {NoStop}%
\bibitem [{\citenamefont {Morgan}(1992)}]{Morgan:1992ge}%
  \BibitemOpen
  \bibfield  {author} {\bibinfo {author} {\bibfnamefont {D.}~\bibnamefont
  {Morgan}},\ }\bibfield  {title} {\bibinfo {title} {{Pole counting and
  resonance classification}},\ }\href
  {https://doi.org/10.1016/0375-9474(92)90550-4} {\bibfield  {journal}
  {\bibinfo  {journal} {Nucl. Phys. A}\ }\textbf {\bibinfo {volume} {543}},\
  \bibinfo {pages} {632} (\bibinfo {year} {1992})}\BibitemShut {NoStop}%
\bibitem [{Note4()}]{Note4}%
  \BibitemOpen
  \bibinfo {note} {The equivalence of the two approaches was first discussed in
  \cite {Baru:2003qq}, generalization to virtual states is provided in \cite
  {Matuschek:2020gqe} and, for a positive effective range to be the case for
  the lattice analyses of $T_{cc}^+$ \cite {Padmanath:2022cvl,Chen:2022vpo}, in
  \cite {Li:2021cue,Albaladejo:2022sux}.}\BibitemShut {Stop}%
\bibitem [{Note5()}]{Note5}%
  \BibitemOpen
  \bibinfo {note} {That zeros in the $T$-matrix invalidate the original
  Weinberg formalism was discussed in \cite
  {Weinberg:1965zz,Baru:2010ww,Kang:2016jxw}.}\BibitemShut {Stop}%
\bibitem [{\citenamefont {Prelovsek}\ and\ \citenamefont
  {Leskovec}(2013)}]{Prelovsek:2013cra}%
  \BibitemOpen
  \bibfield  {author} {\bibinfo {author} {\bibfnamefont {S.}~\bibnamefont
  {Prelovsek}}\ and\ \bibinfo {author} {\bibfnamefont {L.}~\bibnamefont
  {Leskovec}},\ }\bibfield  {title} {\bibinfo {title} {{Evidence for $X(3872)$
  from $DD^*$ scattering on the lattice}},\ }\href
  {https://doi.org/10.1103/PhysRevLett.111.192001} {\bibfield  {journal}
  {\bibinfo  {journal} {Phys. Rev. Lett.}\ }\textbf {\bibinfo {volume} {111}},\
  \bibinfo {pages} {192001} (\bibinfo {year} {2013})},\ \Eprint
  {https://arxiv.org/abs/1307.5172} {arXiv:1307.5172 [hep-lat]} \BibitemShut
  {NoStop}%
\bibitem [{\citenamefont {Braaten}\ and\ \citenamefont
  {Stapleton}(2010)}]{Braaten:2009jke}%
  \BibitemOpen
  \bibfield  {author} {\bibinfo {author} {\bibfnamefont {E.}~\bibnamefont
  {Braaten}}\ and\ \bibinfo {author} {\bibfnamefont {J.}~\bibnamefont
  {Stapleton}},\ }\bibfield  {title} {\bibinfo {title} {{Analysis of $J/\psi
  \pi^+ \pi^-$ and $D^0 \bar D^0 \pi^0$ Decays of the $X(3872)$}},\ }\href
  {https://doi.org/10.1103/PhysRevD.81.014019} {\bibfield  {journal} {\bibinfo
  {journal} {Phys. Rev. D}\ }\textbf {\bibinfo {volume} {81}},\ \bibinfo
  {pages} {014019} (\bibinfo {year} {2010})},\ \Eprint
  {https://arxiv.org/abs/0907.3167} {arXiv:0907.3167 [hep-ph]} \BibitemShut
  {NoStop}%
\bibitem [{\citenamefont {Baru}\ \emph {et~al.}(2022)\citenamefont {Baru},
  \citenamefont {Dong}, \citenamefont {Du}, \citenamefont {Filin},
  \citenamefont {Guo}, \citenamefont {Hanhart}, \citenamefont {Nefediev},
  \citenamefont {Nieves},\ and\ \citenamefont {Wang}}]{Baru:2021ldu}%
  \BibitemOpen
  \bibfield  {author} {\bibinfo {author} {\bibfnamefont {V.}~\bibnamefont
  {Baru}}, \bibinfo {author} {\bibfnamefont {X.-K.}\ \bibnamefont {Dong}},
  \bibinfo {author} {\bibfnamefont {M.-L.}\ \bibnamefont {Du}}, \bibinfo
  {author} {\bibfnamefont {A.}~\bibnamefont {Filin}}, \bibinfo {author}
  {\bibfnamefont {F.-K.}\ \bibnamefont {Guo}}, \bibinfo {author} {\bibfnamefont
  {C.}~\bibnamefont {Hanhart}}, \bibinfo {author} {\bibfnamefont
  {A.}~\bibnamefont {Nefediev}}, \bibinfo {author} {\bibfnamefont
  {J.}~\bibnamefont {Nieves}},\ and\ \bibinfo {author} {\bibfnamefont
  {Q.}~\bibnamefont {Wang}},\ }\bibfield  {title} {\bibinfo {title} {{Effective
  range expansion for narrow near-threshold resonances}},\ }\href
  {https://doi.org/10.1016/j.physletb.2022.137290} {\bibfield  {journal}
  {\bibinfo  {journal} {Phys. Lett. B}\ }\textbf {\bibinfo {volume} {833}},\
  \bibinfo {pages} {137290} (\bibinfo {year} {2022})},\ \Eprint
  {https://arxiv.org/abs/2110.07484} {arXiv:2110.07484 [hep-ph]} \BibitemShut
  {NoStop}%
\bibitem [{\citenamefont {Baru}\ \emph {et~al.}(2015)\citenamefont {Baru},
  \citenamefont {Epelbaum}, \citenamefont {Filin},\ and\ \citenamefont
  {Gegelia}}]{Baru:2015ira}%
  \BibitemOpen
  \bibfield  {author} {\bibinfo {author} {\bibfnamefont {V.}~\bibnamefont
  {Baru}}, \bibinfo {author} {\bibfnamefont {E.}~\bibnamefont {Epelbaum}},
  \bibinfo {author} {\bibfnamefont {A.~A.}\ \bibnamefont {Filin}},\ and\
  \bibinfo {author} {\bibfnamefont {J.}~\bibnamefont {Gegelia}},\ }\bibfield
  {title} {\bibinfo {title} {{Low-energy theorems for nucleon-nucleon
  scattering at unphysical pion masses}},\ }\href
  {https://doi.org/10.1103/PhysRevC.92.014001} {\bibfield  {journal} {\bibinfo
  {journal} {Phys. Rev. C}\ }\textbf {\bibinfo {volume} {92}},\ \bibinfo
  {pages} {014001} (\bibinfo {year} {2015})},\ \Eprint
  {https://arxiv.org/abs/1504.07852} {arXiv:1504.07852 [nucl-th]} \BibitemShut
  {NoStop}%
\bibitem [{\citenamefont {Baru}\ \emph {et~al.}(2016)\citenamefont {Baru},
  \citenamefont {Epelbaum},\ and\ \citenamefont {Filin}}]{Baru:2016evv}%
  \BibitemOpen
  \bibfield  {author} {\bibinfo {author} {\bibfnamefont {V.}~\bibnamefont
  {Baru}}, \bibinfo {author} {\bibfnamefont {E.}~\bibnamefont {Epelbaum}},\
  and\ \bibinfo {author} {\bibfnamefont {A.~A.}\ \bibnamefont {Filin}},\
  }\bibfield  {title} {\bibinfo {title} {{Low-energy theorems for
  nucleon-nucleon scattering at $M_\pi=450$ MeV}},\ }\href
  {https://doi.org/10.1103/PhysRevC.94.014001} {\bibfield  {journal} {\bibinfo
  {journal} {Phys. Rev. C}\ }\textbf {\bibinfo {volume} {94}},\ \bibinfo
  {pages} {014001} (\bibinfo {year} {2016})},\ \Eprint
  {https://arxiv.org/abs/1604.02551} {arXiv:1604.02551 [nucl-th]} \BibitemShut
  {NoStop}%
\bibitem [{\citenamefont {Baru}\ \emph {et~al.}(2004)\citenamefont {Baru},
  \citenamefont {Haidenbauer}, \citenamefont {Hanhart}, \citenamefont
  {Kalashnikova},\ and\ \citenamefont {Kudryavtsev}}]{Baru:2003qq}%
  \BibitemOpen
  \bibfield  {author} {\bibinfo {author} {\bibfnamefont {V.}~\bibnamefont
  {Baru}}, \bibinfo {author} {\bibfnamefont {J.}~\bibnamefont {Haidenbauer}},
  \bibinfo {author} {\bibfnamefont {C.}~\bibnamefont {Hanhart}}, \bibinfo
  {author} {\bibfnamefont {Y.}~\bibnamefont {Kalashnikova}},\ and\ \bibinfo
  {author} {\bibfnamefont {A.~E.}\ \bibnamefont {Kudryavtsev}},\ }\bibfield
  {title} {\bibinfo {title} {{Evidence that the $a_0(980)$ and $f_0(980)$ are
  not elementary particles}},\ }\href
  {https://doi.org/10.1016/j.physletb.2004.01.088} {\bibfield  {journal}
  {\bibinfo  {journal} {Phys. Lett. B}\ }\textbf {\bibinfo {volume} {586}},\
  \bibinfo {pages} {53} (\bibinfo {year} {2004})},\ \Eprint
  {https://arxiv.org/abs/hep-ph/0308129} {arXiv:hep-ph/0308129} \BibitemShut
  {NoStop}%
\bibitem [{\citenamefont {Matuschek}\ \emph {et~al.}(2021)\citenamefont
  {Matuschek}, \citenamefont {Baru}, \citenamefont {Guo},\ and\ \citenamefont
  {Hanhart}}]{Matuschek:2020gqe}%
  \BibitemOpen
  \bibfield  {author} {\bibinfo {author} {\bibfnamefont {I.}~\bibnamefont
  {Matuschek}}, \bibinfo {author} {\bibfnamefont {V.}~\bibnamefont {Baru}},
  \bibinfo {author} {\bibfnamefont {F.-K.}\ \bibnamefont {Guo}},\ and\ \bibinfo
  {author} {\bibfnamefont {C.}~\bibnamefont {Hanhart}},\ }\bibfield  {title}
  {\bibinfo {title} {{On the nature of near-threshold bound and virtual
  states}},\ }\href {https://doi.org/10.1140/epja/s10050-021-00413-y}
  {\bibfield  {journal} {\bibinfo  {journal} {Eur. Phys. J. A}\ }\textbf
  {\bibinfo {volume} {57}},\ \bibinfo {pages} {101} (\bibinfo {year} {2021})},\
  \Eprint {https://arxiv.org/abs/2007.05329} {arXiv:2007.05329 [hep-ph]}
  \BibitemShut {NoStop}%
\bibitem [{\citenamefont {Li}\ \emph {et~al.}(2022)\citenamefont {Li},
  \citenamefont {Guo}, \citenamefont {Pang},\ and\ \citenamefont
  {Wu}}]{Li:2021cue}%
  \BibitemOpen
  \bibfield  {author} {\bibinfo {author} {\bibfnamefont {Y.}~\bibnamefont
  {Li}}, \bibinfo {author} {\bibfnamefont {F.-K.}\ \bibnamefont {Guo}},
  \bibinfo {author} {\bibfnamefont {J.-Y.}\ \bibnamefont {Pang}},\ and\
  \bibinfo {author} {\bibfnamefont {J.-J.}\ \bibnamefont {Wu}},\ }\bibfield
  {title} {\bibinfo {title} {{Generalization of Weinberg\textquoteright{}s
  compositeness relations}},\ }\href
  {https://doi.org/10.1103/PhysRevD.105.L071502} {\bibfield  {journal}
  {\bibinfo  {journal} {Phys. Rev. D}\ }\textbf {\bibinfo {volume} {105}},\
  \bibinfo {pages} {L071502} (\bibinfo {year} {2022})},\ \Eprint
  {https://arxiv.org/abs/2110.02766} {arXiv:2110.02766 [hep-ph]} \BibitemShut
  {NoStop}%
\bibitem [{\citenamefont {Albaladejo}\ and\ \citenamefont
  {Nieves}(2022)}]{Albaladejo:2022sux}%
  \BibitemOpen
  \bibfield  {author} {\bibinfo {author} {\bibfnamefont {M.}~\bibnamefont
  {Albaladejo}}\ and\ \bibinfo {author} {\bibfnamefont {J.}~\bibnamefont
  {Nieves}},\ }\bibfield  {title} {\bibinfo {title} {{Compositeness of S-wave
  weakly-bound states from next-to-leading order Weinberg\textquoteright{}s
  relations}},\ }\href {https://doi.org/10.1140/epjc/s10052-022-10695-1}
  {\bibfield  {journal} {\bibinfo  {journal} {Eur. Phys. J. C}\ }\textbf
  {\bibinfo {volume} {82}},\ \bibinfo {pages} {724} (\bibinfo {year} {2022})},\
  \Eprint {https://arxiv.org/abs/2203.04864} {arXiv:2203.04864 [hep-ph]}
  \BibitemShut {NoStop}%
\bibitem [{\citenamefont {Baru}\ \emph {et~al.}(2010)\citenamefont {Baru},
  \citenamefont {Hanhart}, \citenamefont {Kalashnikova}, \citenamefont
  {Kudryavtsev},\ and\ \citenamefont {Nefediev}}]{Baru:2010ww}%
  \BibitemOpen
  \bibfield  {author} {\bibinfo {author} {\bibfnamefont {V.}~\bibnamefont
  {Baru}}, \bibinfo {author} {\bibfnamefont {C.}~\bibnamefont {Hanhart}},
  \bibinfo {author} {\bibfnamefont {Y.~S.}\ \bibnamefont {Kalashnikova}},
  \bibinfo {author} {\bibfnamefont {A.~E.}\ \bibnamefont {Kudryavtsev}},\ and\
  \bibinfo {author} {\bibfnamefont {A.~V.}\ \bibnamefont {Nefediev}},\
  }\bibfield  {title} {\bibinfo {title} {{Interplay of quark and meson degrees
  of freedom in a near-threshold resonance}},\ }\href
  {https://doi.org/10.1140/epja/i2010-10929-7} {\bibfield  {journal} {\bibinfo
  {journal} {Eur. Phys. J. A}\ }\textbf {\bibinfo {volume} {44}},\ \bibinfo
  {pages} {93} (\bibinfo {year} {2010})},\ \Eprint
  {https://arxiv.org/abs/1001.0369} {arXiv:1001.0369 [hep-ph]} \BibitemShut
  {NoStop}%
\bibitem [{\citenamefont {Kang}\ and\ \citenamefont
  {Oller}(2017)}]{Kang:2016jxw}%
  \BibitemOpen
  \bibfield  {author} {\bibinfo {author} {\bibfnamefont {X.-W.}\ \bibnamefont
  {Kang}}\ and\ \bibinfo {author} {\bibfnamefont {J.~A.}\ \bibnamefont
  {Oller}},\ }\bibfield  {title} {\bibinfo {title} {{Different pole structures
  in line shapes of the $X(3872)$}},\ }\href
  {https://doi.org/10.1140/epjc/s10052-017-4961-z} {\bibfield  {journal}
  {\bibinfo  {journal} {Eur. Phys. J. C}\ }\textbf {\bibinfo {volume} {77}},\
  \bibinfo {pages} {399} (\bibinfo {year} {2017})},\ \Eprint
  {https://arxiv.org/abs/1612.08420} {arXiv:1612.08420 [hep-ph]} \BibitemShut
  {NoStop}%
\end{thebibliography}%

\begin{appendix}

\section{Theoretical framework and fitting procedure} \label{app:all}

\subsection{Left-hand cuts from pion exchanges}

\begin{figure}[t]
\begin{center}
\includegraphics[width=0.15\textwidth]{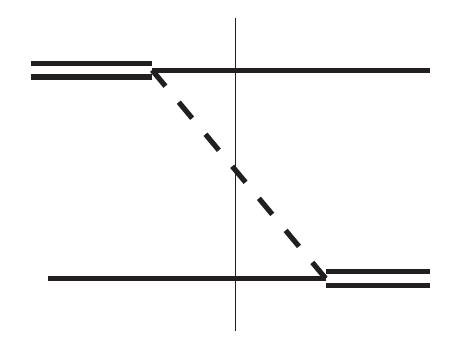}\hspace*{0.05\textwidth}
\includegraphics[width=0.15\textwidth]{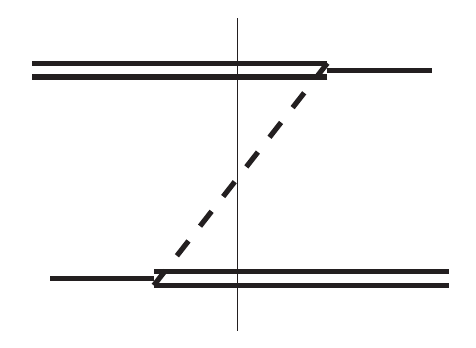}
\caption{The two contributions to the OPE interaction between the $D^*$ (double solid line) and $D$ (single solid line) mesons; the dashed line corresponds to the pion. The thin vertical line picks up the relevant intermediate state: $DD\pi$ and $D^*D^*\pi$ for the left and right plot, respectively.}
\label{fig:TOPT}
\end{center}
\end{figure}

Here we briefly outline the relation between the pion propagator in the Feynman technique and in the time-ordered perturbation theory (TOPT), for details see, {\it e.g.}, \cite{Schweber:1961zz,Baru:2019ndr}.

We start from the Feynman propagator,
\be
D^\pi(q)=\frac{1}{q_\mu q^\mu-m_\pi^2}=\frac{1}{q_0^2-\omega_\pi^2(q^2)},
\ee
where $\omega_\pi(q^2)=\sqrt{q^2+m_\pi^2}$, and rewrite it identically in the form
\be
D^\pi(q)=\frac{1}{{2}\omega_\pi(q^2)}\left(\frac{1}{q_0-\omega_\pi(q^2)}-\frac{1}{q_0+\omega_\pi(q^2)}\right).
\label{Dpiq}
\ee
 
The derivation of the integral equations yields that the energy transfer $q_0$ in the first and second terms in the parentheses can be substituted as 
$q_0=E-E_D(k^2)-E_{D}(k^{\prime 2})$ and $q_0=-E+E_{D^*}(k^2)+E_{D^*}(k^{\prime 2})$, respectively, with ${\bm k}$ and ${\bm k}'$ for the momenta in the initial and final state, respectively. Then the pion propagator (\ref{Dpiq}) finally reads
\begin{eqnarray}
D^\pi(q)&=&\frac{1}{{2}\omega_\pi(q^2)}
\left[\frac{1}{E-E_D(k^2)-E_D(k^{\prime 2})-\omega_\pi(q^2)}\right.\nonumber\\
&&+\left.\frac{1}{E-E_{D^*}(k^2)-E_{D^*}(k^{\prime 2})-\omega_\pi(q^2)}\right], \label{Dpiq2}
\end{eqnarray}
which can be recognized as the TOPT form of the propagator, with the two terms in square brackets for the two time orderings corresponding to the TOPT diagrams shown in Fig.~\ref{fig:TOPT}.

The lhc branch points from the OPE for the $DD^* $ scattering can be found as the end-point singularities of the $DD\pi$ TOPT Green's function in the on-shell kinematics, 
\begin{eqnarray}
&&E - E_{D}(k^2) -E_{D}(k^{\prime 2}) - \omega_\pi(q^2)\nonumber \xrightarrow[\text{on shell: } k=k' =p]{\cos\theta =\pm1} \nonumber\\ [-1mm]
\\[-1mm]
&&
E_{D*}(p^2) - E_{D}(p^2) - \omega_\pi(2p^2(1-\cos\theta))\big|_{\cos\theta=\pm 1}=0, \nonumber
\end{eqnarray}
which gives for the lhc branch point near the two-body threshold ($\cos\theta=-1$)
\begin{eqnarray}
(p_\lhc^{1\pi})^2&=& \frac14\Bigl[(\Delta M)^2-m_\pi^2\Bigr]\frac{(M_D+M_{D^*})^2-m_\pi^2}{2M_D^2+2M_{D^*}^2-m_\pi^2}\nonumber \\ [-2mm]
\label{Eq:lhc1pi}\\[-2mm]
& =& \frac14\Bigl[(\Delta M)^2-m_\pi^2\Bigr] \left [1 + {\cal O}\left(\frac{(\Delta M)^2}{M_D^2}\right)\right]. \nonumber
\end{eqnarray}
Since $\Delta M\ll M_D$, the relativistic correction is negligible for the problem at hand
and we arrive at equation \eqref{plhc} in the main text.
The position of the
more distant branch point corresponding to $\cos\theta=+1$
depends on a particular relation between the masses of the particles involved. In the kinematics discussed in this paper, we have 
\be
\sqrt{M_D^2+m_\pi^2}\leqslant M_{D^*}\leqslant M_D+m_\pi,
\ee
which gives
\begin{eqnarray}
\hspace*{-0.7cm}(\tilde p_\lhc^{1\pi})^2&=& \frac14\Bigl[(\Delta M)^2-m_\pi^2\Bigr] \frac{(M_D+M_{D^*})^2-m_\pi^2}{m_\pi^2}\nonumber \\[-2mm] 
\label{Eq:lhc1pidist}\\[-2mm]
\hspace*{-1.2cm}& =& \frac14\Bigl[(\Delta M)^2-m_\pi^2\Bigr] \frac{4 M_D^2 }{m_\pi^2} \left [1 + {\cal O}\left(\frac{\Delta M}{M_D}\right)\right]. \nonumber
\end{eqnarray}

It follows from \eqref{Eq:lhc1pi} and \eqref{Eq:lhc1pidist} that both lhc branch points move towards the two-body threshold when $m_\pi$ decreases, so in the limit $m_\pi=\Delta M$ the lhc disappears. We would like to highlight that the $D^*D^*\pi$ TOPT Green's function generates only a remote lhc for the $DD^*$ scattering that is irrelevant to the discussion and is thus disregarded.

The next in importance lhc comes from the two-pion exchange (TPE) contributions. A rough estimate of its near-threshold branch point can be done if one employs that 
$(\Delta M)^2/m_\pi^2\ll 1$ for the lattice settings in \cite{Padmanath:2022cvl} to neglect the difference in the mass squared relative to the pion mass squared, which gives $(p_\lhc^{2\pi})^2\approx -m_\pi^2$. 
This implies that although the TPE contributions may introduce certain corrections, the relatively large distance of $(p_\lhc^{2\pi})^2$ from the threshold suggests that there should be no qualitative changes in the extracted poles. Furthermore, 
it is argued in \cite{Baru:2015ira,Baru:2016evv} that
a significant portion of the multi-pion-exchange cut contributions is already accounted for through iterations of the OPE. 

\subsection{\texorpdfstring{$\bm T$}{T}-matrix}
\label{app:framework}

In the isospin limit, the contact potential up to next-to-leading order in the nonrelativistic expansion for the $S$-wave $DD^*$ isoscalar ($I=0$) state reads
\bea
V_C^{I=0}(k,k') = c_0+c_2(k^2+k'^2),
\eea
where $c_0$ and $c_2$ are the low-energy constants and $k$ $(k')$ is the 3-momentum of the incoming (outgoing) particle in the the center-of-mass (c.m.) frame. In the framework of the time-ordered-perturbation theory, the one-pion exchange (OPE) potential is written as
\bea
V_\text{OPE}^{I=0}(E,k,k')& =& \frac{g^2}{8f_\pi^2}\int_{-1}^1 dz D^\pi(E,k,k',z)\nonumber\\
&& \hspace*{0.04\textwidth}\times (k^2+k'^2-2kk'z),
\label{OPEfull}
\eea
where, according to (\ref{Dpiq2}),
$$
D^\pi(E,k,k',z)=-\frac{D_1(E,k,k',z)+D_2(E,k,k',z)}{2\omega_\pi(q^2)},
$$
with
\bea
D_1^{-1}(E,k,k',z) & = &2M_D+\frac{k^2+k'^2}{2M_D}+\omega_\pi(q^2)-E-i\epsilon, \nonumber\\
D_2^{-1}(E,k,k',z) & = &2M_{D^*}+\frac{k^2+k'^2}{2M_{D^*}}+\omega_\pi(q^2)-E-i\epsilon,\nonumber
\eea
and $\omega_\pi(q^2)=\sqrt{m_\pi^2+k^2+k'^2-2kk'z}$, that is, the pion is treated fully relativistically while the $D^{(*)}$ mesons are nonrelativistic.

The $T$-matrix (amplitude) is obtained as a solution of the Lippmann-Schwinger equation
\bea
\label{eq:lse}
T(E,k,k') &=& V(E,k,k') \\
& &- \int\frac{\text{d}^3{\bm q}}{(2\pi)^3}V(E,k,q)G(E,q)T(E,q,k'),\nonumber
\eea
where the potential is a sum of the contact term and OPE, 
\be
V(E,k,k')=V_C(k,k')+V_\text{OPE}(E,k,k').
\ee

In order to render the integral in \eqref{eq:lse} well defined, we set a sharp cutoff $\Lambda$. To arrive at the results in Fig.~\ref{fig:fit}, $\Lambda$ is chosen to be 500 MeV.
The $DD^*$ propagator is expressed as 
\bea
G(E,q) = \left[M_{D^*}+M_D+\frac{q^2}{2\mu}-E-\frac{i}{2}\Gamma(E,q)\right]^{-1},~~
\eea
where $\mu = M_D M_{D^*}/(M_D+M_{D^*})$ is the reduced mass,
$$
\Gamma(E,q) = \frac{g^2M_D}{8\pi f_\pi^2 M_{D^*}}\Big[\Sigma(s) - \Sigma_0(s)\theta(M_D+m_\pi-M_{D^*}) \Big],
$$
with $s=[E-M_D-q^2/(2\mu)]^2$, and
\bea
\Sigma(s) = \left[ \frac{\sqrt{\lambda(s,M_D^2,m_\pi^2)}}{2\sqrt{s} }\right]^{3},
\eea
where $\lambda(a,b,c)=a^2+b^2+c^2-2ab-2bc-2ca$ is the K\"all\'en triangle function. Here
\bea
\Sigma_0(s)&=& \Sigma(M_{D^*}^2)\nonumber\\
&&+ 2M_{D^*}\left(E-M_{D^*}-M_D-\frac{q^2}{2\mu}\right)\Sigma'(M_{D^*}^2),\nonumber
\eea
where the first and second terms renormalize the $D^*$ mass and wave function, respectively, if $M_{D^*}<M_D+m_\pi$.

\begin{figure*}[t!]
\begin{center}
\includegraphics[width=0.41\textwidth]{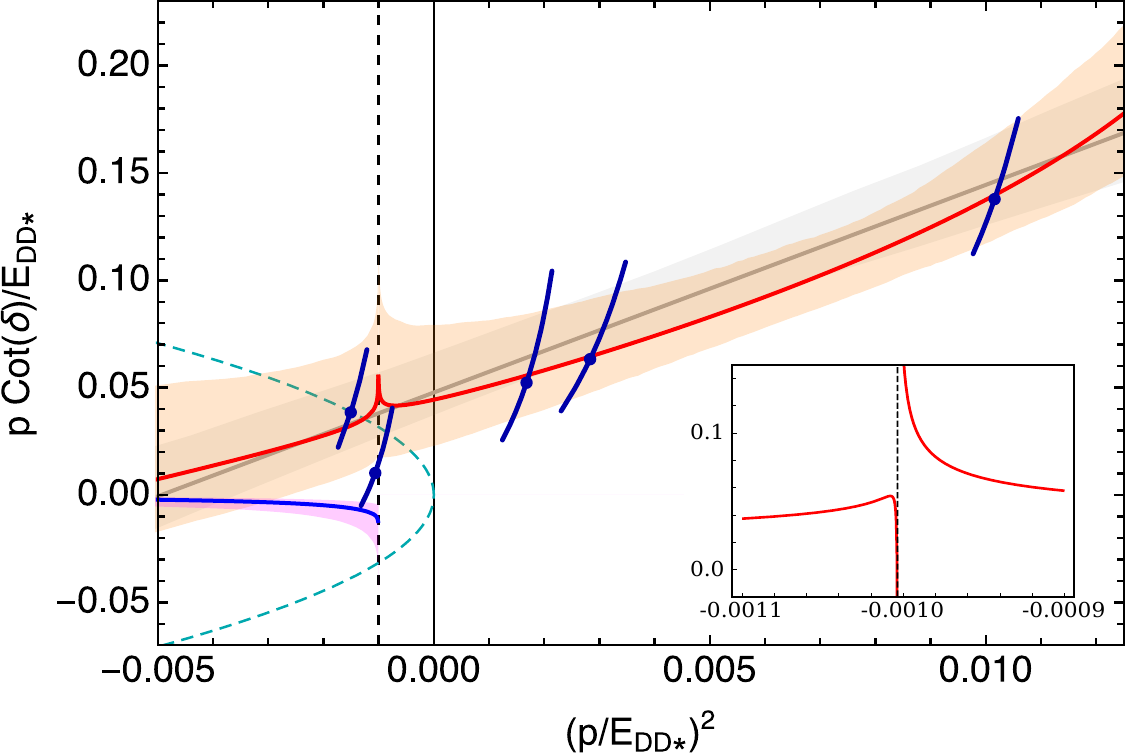}\hspace*{0.05\textwidth}
\includegraphics[width=0.41\textwidth]{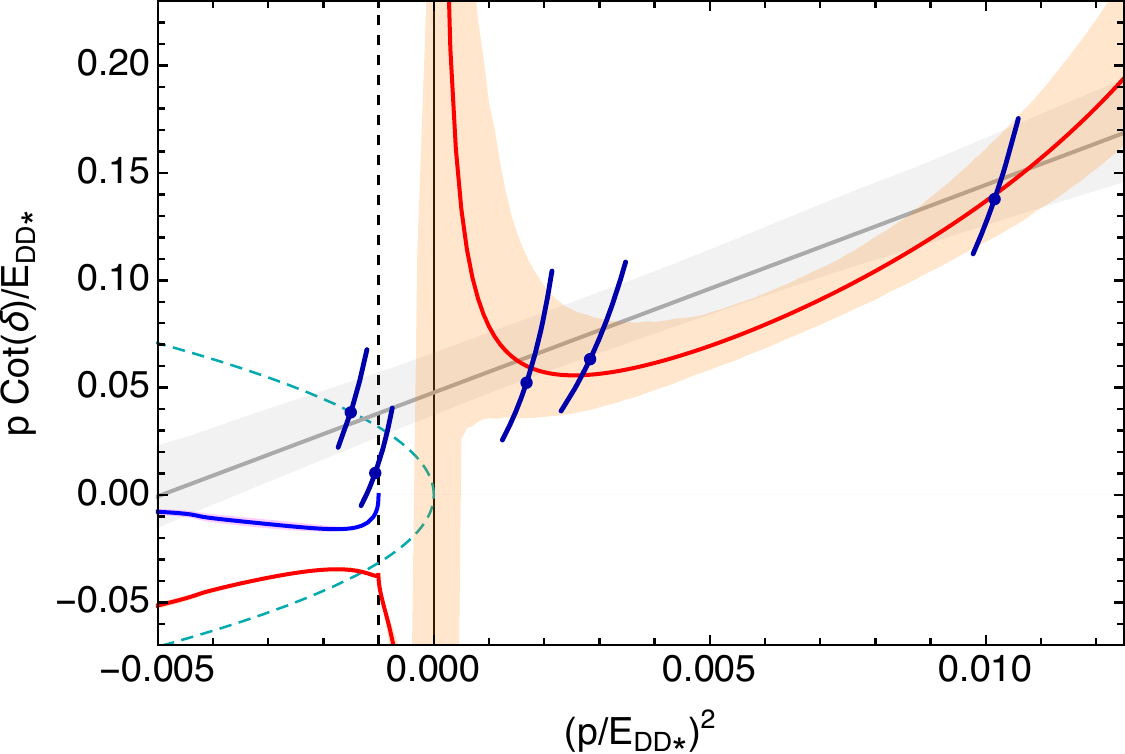}
\caption{Fits for the OPE potential artificially suppressed by a factor of 1/10 (the left plot) or artificially enhanced by a factor of 5 (the right plot). The inset in the left panel magnifies the red curve in the vicinity of the lhc branch point. See the caption of Fig.~\ref{fig:fit} in the main text for further details. {Here only the three data points not affected by the lhc are considered in the fits.}}
\label{fig:fitVpi}
\end{center}
\end{figure*}

\subsection{Chiral extrapolation of \texorpdfstring{$\bm{f_\pi}$}{fpi} and \texorpdfstring{$\bm g$}{g}}\label{app:chiral}

For convenience, we introduce the ratio $\xi=m_\pi/m_\pi^\text{ph}$ with $m_\pi^\text{ph}$ the physical pion mass. For the masses of the $D$ and $D^*$ we stick to the values used in the lattice calculations and quoted explicitly in \cite{Padmanath:2022cvl}. For the pion decay constant $f_\pi$, we resort to the result of the one-loop chiral perturbation theory \cite{Gasser:1983yg} and cast it into the form~\cite{Baru:2013rta}
\bea
f_\pi(\xi)=f_\pi^\text{ph}\left[1+\left(1-\frac{f_0}{f_\pi^\text{ph}}\right)(\xi^2-1)-\frac{(m_\pi^\text{ph})^2}{8\pi^2f_0^2}\xi^2\log\xi \right], \nonumber
\eea
where $f_0 \equiv f_\pi(m_\pi=0)=85$ MeV \cite{Becirevic:2012pf} and $f_\pi^\text{ph} = 92.1~\text{MeV}$.

\begin{figure*}[t!]
\begin{center}
\includegraphics[width=0.42\textwidth]{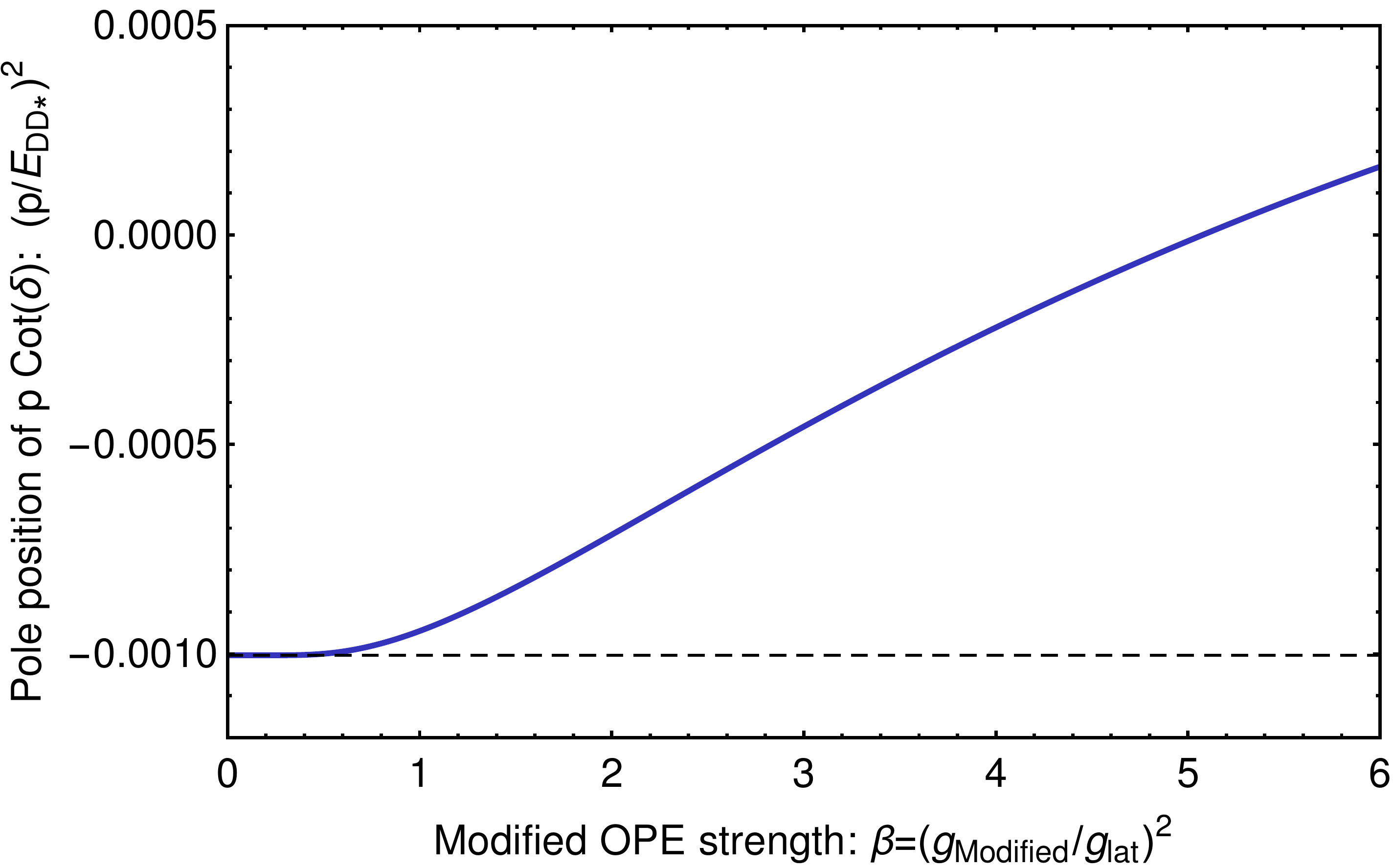}
\hspace*{0.05\textwidth}
\includegraphics[width=0.38\textwidth]{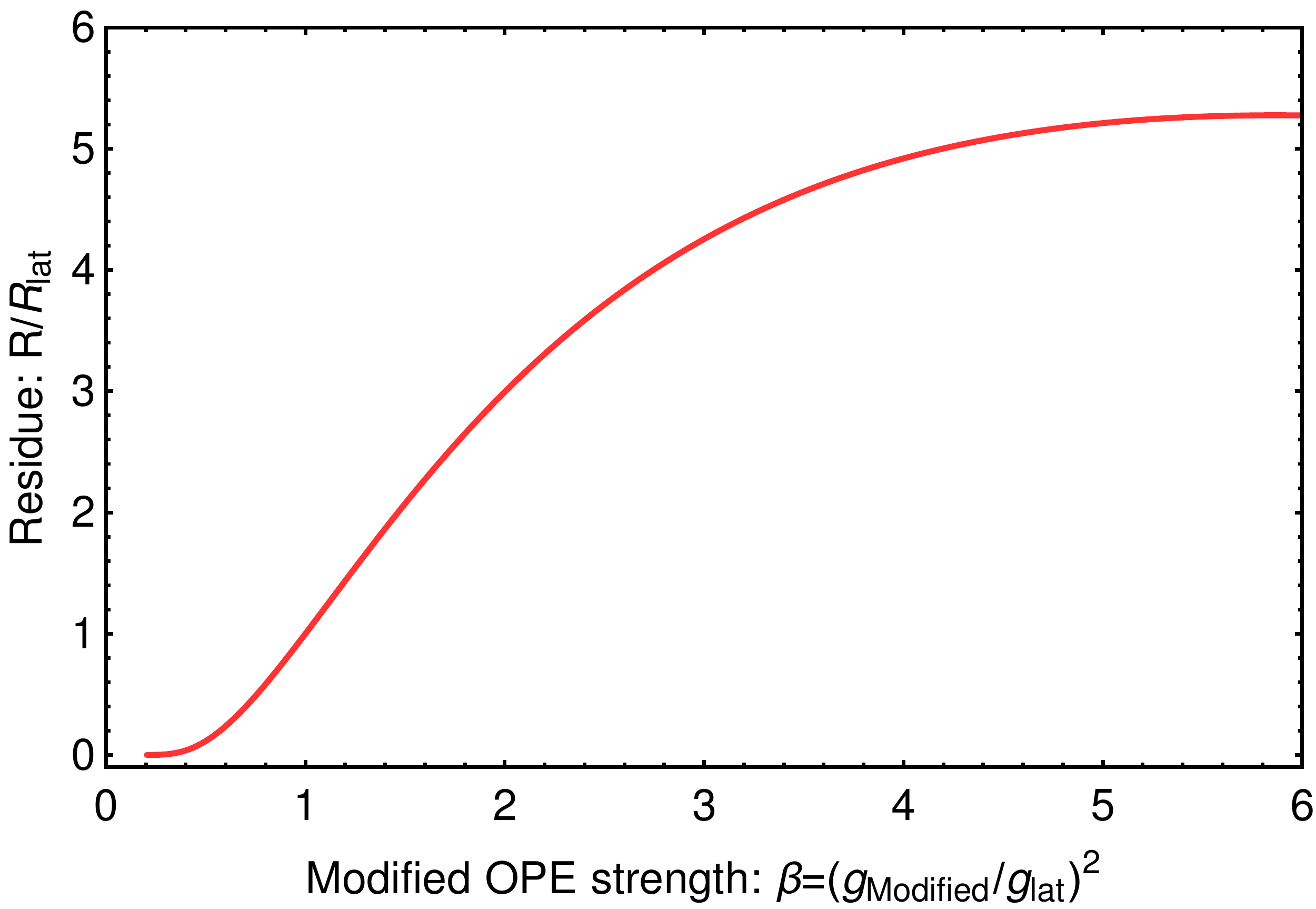}
\caption{The behavior of the pole position (left) and the residue (right) of $p\cot\delta$ with the variation of the OPE strength while keeping the pion mass fixed. Only the central values are shown. The horizontal dashed line in the left panel indicates the lhc. The data points used in the fits for the artificially suppressed/enhanced OPE potential are the same as Fig.~\ref{fig:fitVpi}.}
\label{fig:polebehavior}
\end{center}
\end{figure*}

For the chiral extrapolation of the $DD^*\pi$ coupling $g$, we employ the lattice results of \cite{Becirevic:2012pf} and express it in terms of the physical value $g^\text{ph}$ and $g_0 \equiv g(m_\pi=0)$ as \cite{Baru:2013rta}
\be
g(\xi) = g^\text{ph}\left[1+C_1 (\xi^2-1) +C_2\xi^2
\log\xi \right], 
\ee
where $g^\text{ph}=0.57$ is determined from the experimentally measured $D^{*+}\to D^0\pi^+$ decay width \cite{Du:2021zzh},
\bea
&&C_1 = 1- \left[ 1 - \frac{1+2g_0^2}{8\pi^2 f_0^2}(m_\pi^\text{ph})^2\log\frac{m_\pi^\text{ph}}{\mu_\text{lat}} + \alpha_\text{lat}(m_\pi^\text{ph})^2\right]^{-1} , \nonumber\\
&& C_2 = -\frac{1+2g_0^2}{8\pi^2 f_0^2} (m_\pi^\text{ph})^2 (1-C_1),\nonumber
\eea
and the parameters take the values \cite{Becirevic:2012pf}
\bea
g_0=0.46, \quad \alpha_\text{lat}=-0.16~\text{GeV}^{-2}, \quad \mu_\text{lat} = 1~\text{GeV}.\nonumber
\eea
Specifically, for $m_\pi=280$ MeV this gives 
\be\label{eq:glat}
g(m_\pi=280\, {\rm MeV}) = 0.65 .
\ee

The variation of the strength of the OPE potential (within approximately 30\%, driven by
the pion mass dependence of both $f_\pi$ and $g$) due to changing $m_\pi$ between its physical and lattice values provides only a subleading effect on the dynamics of the system at hand while the leading effect comes from the position of the lhc branch point, as given in \eqref{plhc}. Specifically, for $m_{\pi}> \Delta M$ the lhc is on the real axis in the energy plane with the lhc branch point being close to the threshold.

\begin{figure*}[t!]
\begin{center}
\includegraphics[width=0.89\textwidth]{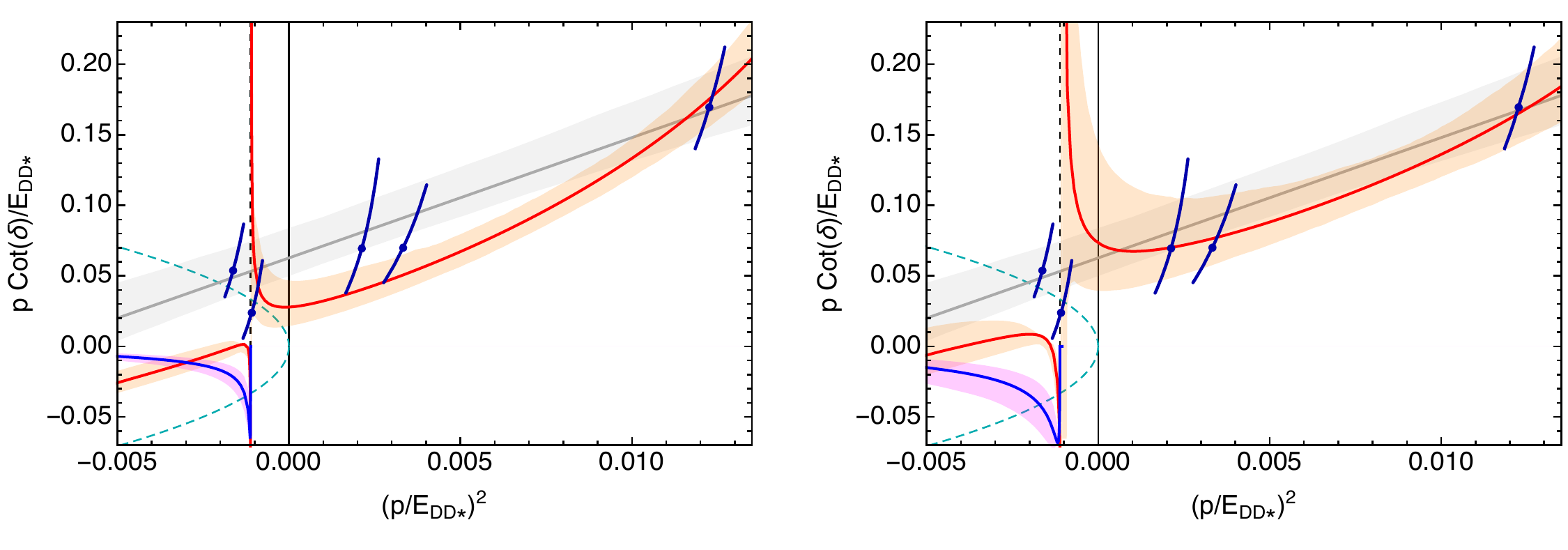}
\caption{Fits for the alternative set of parameters provided in \cite{Padmanath:2022cvl} and quoted in (\ref{params2}). See the caption of Fig.~\ref{fig:fit} in the main text for further details.}
\label{fig:fitsetup2}
\end{center}
\end{figure*}

\subsection{Fitting procedure}\label{app:fitting}

In this appendix we describe the procedure employed to fit $p \cot \delta(p^2)$ extracted from the lattice data in
\cite{Padmanath:2022cvl}
using the model described above which explicitly takes into account the OPE interaction.

In particular, the function $p \cot \delta(p^2)$ is related to the $DD^*$ $S$-wave on-shell scattering $T$-matrix $T(E)$ as
\begin{equation}
p \cot \delta(p^2) = -\frac{2 \pi }{\mu}T^{-1}(E) + i p,
\end{equation}
where
$T(E) \equiv T(M_D+M_{D^*}+p^2/(2\mu), p, p)$ is a solution of the Lippmann-Schwinger equation~(\ref{eq:lse}).

Since the lattice data points for $p \cot \delta(p^2)$ are extracted
using the L\"uscher method, their uncertainties have non-Gaussian shapes, which prevents us from performing a naive $\chi^2$ fit. Instead, 
we employ the probability distribution extracted for each lattice data point from 
\cite{Padmanath:2022cvl}
and generate 1000 sets of quasi-data points according to these distributions.
Each generated set (with 3 or 4 data points depending on the fit) is then fitted with our model by minimizing
$$
\chi^2 = \sum_i \left(p_i \cot \delta{(p_i^2)}\big|_{\text{model}} - p_i \cot \delta{(p_i^2)}\big|_{\text{quasi-data}}\right)^2
$$
as a function of the parameters $c_0$ and $c_2$,
where $i$ enumerates the data points used in the fit---see the main text for the details of the selection procedure of the data points.
The resulting 1000 pairs $\{c_0,\,c_2\}$ are used then to evaluate the most probable values of these low-energy constants, $p \cot \delta(p^2)$, and the corresponding 68\% confidence intervals.
{The correlations between lattice data points are not available in \cite{Padmanath:2022cvl}, so that we do not consider them in our fits. It is sufficient for the illustrative purpose of the significant impact of the OPE potential on the pole extraction as we do not aim at a strict and comprehensive analysis of the lattice data from \cite{Padmanath:2022cvl}, which calls for a well-justified approach taking the lhc into account. }

\subsection{Pionic dynamics for unphysically small and large pion coupling}
\label{app:cdd}

Here we discuss the effect of the pionic dynamics in case of the OPE potential artificially suppressed by a factor of 1/10 (the left plot in Fig.~\ref{fig:fitVpi}) or enhanced by a factor of 5 (the right plot in Fig.~\ref{fig:fitVpi}). From these plots, in agreement with the natural expectations, one can see that, for a weaker pion exchange, the fit approaches the ERE linear formula. On the contrary, for the enhanced OPE interaction, the singularity of $p\cot\delta(p^2)$ moves closer to the two-body threshold, thus limiting the range of convergence of the ERE
even stronger than the lhc. Therefore, an approximate coincidence of the pole in $p\cot\delta$ and lhc, that takes place for the physical value of the pion coupling, comes as a result of fitting the particular data points taken from \cite{Padmanath:2022cvl}.

{In order to show the essential effect of the OPE on the pole of $p\cot\delta$, we demonstrate how the pole moves if we vary the OPE strength while keeping the pion mass fixed---see the left panel in Fig.~\ref{fig:polebehavior}. For the lattice OPE strength from \eqref{eq:glat}, the pole of $p\cot\delta$ appears just above the lhc. Then, if the pion coupling (labeled as $g_\text{Modified}$) is artificially decreased, the pole gradually approaches the lhc. 
Finally, as the OPE strength vanishes ($\beta=(g_\text{Modified}/g_\text{lat})^2\to 0$), the residue at the pole also vanishes (see the right panel in Fig.~\ref{fig:polebehavior}), ensuring that OPE is essential for producing the pole in $p\cot\delta$. 
}

\section{Analysis of the lattice data for the smaller charm quark masses} 
\label{app:fit2}

In this appendix we provide the fit results for the lattice data from \cite{Padmanath:2022cvl}
corresponding to the smaller charm quark masses, namely 
\be
M_D=1762~\mbox{MeV},\quad M_{D^*}=1898~\mbox{MeV},
\label{params2}
\ee
while the pion mass remains the same as before, $m_\pi=280$~MeV. The obtained 3- and 4-point fits are shown in Fig.~\ref{fig:fitsetup2}. In these settings, the lhc position corresponds to ${(p/E_{DD^*})}^2 = -0.0011$. As stated in the main text, these fits demonstrate essentially the same behaviour as those shown in Fig.~\ref{fig:fit} and as such lead to the same conclusions.

\end{appendix}

\end{document}